\documentclass{emulateapj}
\usepackage{graphicx}
\usepackage[bookmarks=true,breaklinks=true]{hyperref}
\usepackage{amsmath}
\usepackage{lineno}

\begin{document}
\newcommand{\degree}{^{\circ}}

\title{Long-Term Radio Timing Observations of the Transition Millisecond Pulsar PSR~J1023+0038}
\shorttitle{Timing J1023}
\shortauthors{Archibald et al.}
\author{Anne M. Archibald\altaffilmark{1,2,*}}
\author{Victoria M. Kaspi\altaffilmark{1}}
\author{Jason W. T. Hessels\altaffilmark{2,3}}
\author{Ben Stappers\altaffilmark{4}}
\author{Gemma Janssen\altaffilmark{4,2}}
\author{Andrew G. Lyne\altaffilmark{4}}

\altaffiltext{*}{Corresponding author; email should be addressed to \protect \email{archibald@astron.nl}}
\altaffiltext{1}{Department of Physics, McGill University, 3600 University St., Montreal, QC H3A 2T8, Canada}
\altaffiltext{2}{ASTRON, the Netherlands Institute for Radio Astronomy, Postbus 2, 7990 AA, Dwingeloo, The Netherlands}
\altaffiltext{3}{Astronomical Institute ``Anton Pannekoek,'' University of Amsterdam, Science Park 904, 1098 XH Amsterdam, The Netherlands}
\altaffiltext{4}{Jodrell Bank Centre for Astrophysics, School of Physics and Astronomy, The University of Manchester, M13 9PL, UK}

\begin{abstract}
    The radio millisecond pulsar PSR J1023+0038 exhibits complex timing and eclipse behavior. Here we analyze four years' worth of radio monitoring observations of this object. We obtain a long-term timing solution, albeit with large residual timing errors as a result of apparent orbital period variations. We also observe variable eclipses when the companion passes near our line of sight, excess dispersion measure near the eclipses and at random orbital phases, and short-term disappearances of signal at random orbital phases. We interpret the eclipses as possibly due to material in the companion's magnetosphere supported by magnetic pressure, and the orbital period variations as possibly due to a gravitational quadrupole coupling mechanism. Both of these mechanisms would be the result of magnetic activity in the companion, in conflict with evolutionary models that predict it should be fully convective and hence non-magnetic. We also use our timing data to test for orbital and rotational modulation of the system's $\gamma$-ray emission, finding no evidence for orbital modulation and $3.7\sigma$ evidence for modulation at the pulsar period. The energetics of the system make it plausible that the $\gamma$-ray emission we observe is entirely from the millisecond pulsar itself, but it seems unlikely for these $\gamma$-rays to provide the irradiation of the companion, which we attribute instead to X-ray heating from a shock powered by a particle wind. 
\end{abstract}

\maketitle

\section{Introduction}
The eclipsing millisecond radio pulsar PSR J1023+0038\footnote{Also known as FIRST~J102347.67+003841.2.} (hereafter J1023) has a nearly unique history: it likely had an accretion disc for approximately one year circa 2001 \citep{asr+09}. In 2007 it was discovered as a radio pulsar with a spin period of $1.7~\text{ms}$ and an orbital period of $4.8$ hours, making it a transition object between low-mass X-ray binaries (LMXBs) and millisecond pulsars (MSPs). Its companion has a mass of $0.2~M_\sun$ and appears to be a G star, mildly irradiated by the pulsar. Based on radio observations, \citet{asr+09} also reported eclipses, dispersion measure (DM) variations, and orbital period variations, but the short span of observations then available prevented an accurate description. 

More than thirty eclipsing MSPs are now known, including at least $13$ in the Galactic field \citep{robe11}. \citet{fst88} discovered the first of the ``black widow'' systems, PSR~B1957+20, in which the MSP appears to be ablating a low-mass degenerate companion. A second class has recently emerged, of which J1023 is the holotype, the ``redbacks'' \citep{robe11}, comprising several recently discovered systems in the Galactic field and several previously known systems in globular clusters. The redbacks have non-degenerate, main-sequence-like, companions that are more massive than those of the black widows. Both types of system exhibit orbital period variations \citep[e.g.][]{dlk+01} and various types of radio phenomenology. In many cases it is clear that the companion's Roche lobe never crosses the line of sight, so that the eclipses must be due to ionized material that is not gravitationally bound \citep[e.g.][]{sbl+01}. In light of J1023's history, the loose material in the redbacks may be related to their transition from LMXBs to radio MSPs.

The link between radio MSPs and LMXBs has recently been strengthened by observations by \citet{pfb+13} of the radio MSP PSR~J1824$-$2452I in the globular cluster M28. Radio timing observations of this MSP are consistent with it having a main-sequence companion, perhaps qualifying it as a redback. The LMXB IGR~J18245$-$2452 was recently observed to exhibit coherent X-ray pulsations matching the ephemeris of the radio MSP. The identification of these two sources shows that an object can swing back and forth between a radio MSP and a LMXB. In fact, it appears that the transition from LMXB to MSP can happen in no more than a few weeks \citep{phb+13}. This object promises to provide invaluable insights into the transition process, but it is much more distant and consequently fainter than J1023, so our own observations should complement observations of this new system. 
In its radio pulsar state, J1023 has been the target of intensive multi-wavelength studies. \citet{akb+10} and \citet{bah+11} found that it exhibits variable X-ray emission apparently modulated at both rotational and orbital periods. \citet{thh+10} noted that J1023 is associated with a $\gamma$-ray source, and indeed a source at the position of J1023 appears in the second \emph{Fermi} catalog \citep[2FGL;][]{naa+12}. \citet{wwm13} searched for evidence in the infrared of a circumstellar dust cloud due to ejected disk material, but found none. \citet{das+12} carried out a very-long-baseline radio interferometry campaign to measure the position, proper motion, and parallax of J1023, measuring its distance to be $1368^{+42}_{-39}$ pc and estimating the pulsar mass as $1.71\pm 0.16\ M_\Sun$. 

In this paper we use several years of radio timing observations to describe the pulsar's longer-term behavior. In Section~\ref{sec:observations} we summarize the observation systems and cadence; in Section~\ref{sec:timing} we discuss long-term timing solutions for J1023; in Section~\ref{sec:radiophenomenology} we describe the eclipses and other phenomena; in Section~\ref{sec:gamma} we look at the $\gamma$-ray behavior in light of our timing solution; and in Section~\ref{sec:discussion} we discuss the implications our observations have for understanding the system. 

\section{Observations}
\label{sec:observations}

\begin{figure}
    \plotone{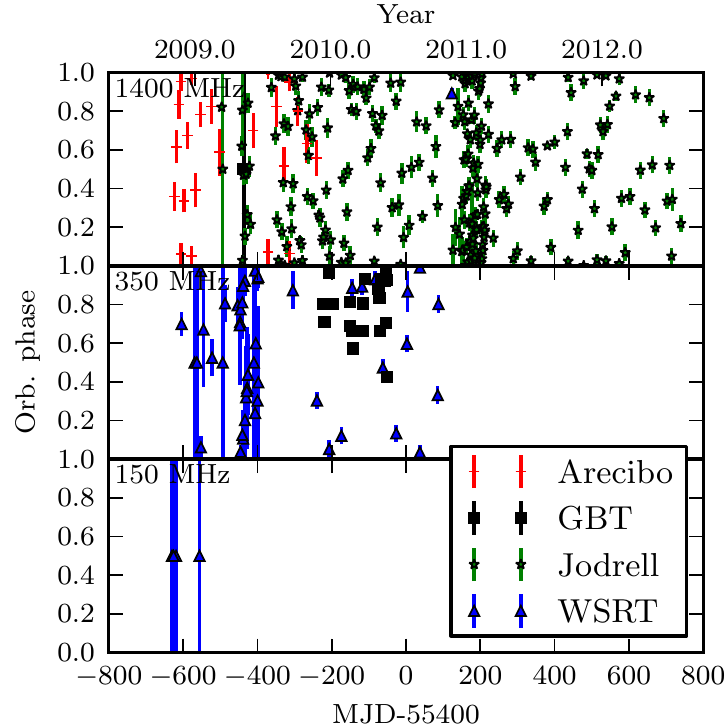}
    \caption{\label{fig:monitoring}Our radio observations of J1023, showing the range of orbital phases covered and the date. The top panel shows 1400-MHz observations, middle panel shows 350-MHz observations, and bottom panel shows 150-MHz observations. Colors and markers indicate observatory, with red crosses denoting Arecibo, green stars denoting Jodrell Bank, black squares denoting Green Bank, and blue triangles denoting Westerbork. Observations longer than a full orbit are marked here as top-to-bottom bars.}
\end{figure}

We have carried out radio timing observations of J1023 spanning four years with four different observing systems, described below. The observations are summarized in Figure~\ref{fig:monitoring}.

Observations at the Arecibo Observatory in Puerto Rico were carried out with the 300-m William E. Gordon radio telescope. We used the L-band wide receiver, which makes available four independent 100-MHz intermediate-frequency channels (IFs). We fed one of these IFs into the ASP coherent dedispersion backend \citep{demo07,ferd08}, producing online-folded data in PSRFITS format spanning 1384 to 1430~MHz. We fed the other three IFs into three WAPPs \citep{dsh00}, which produced search-mode\footnote{Where possible we obtained data in search mode in order to retain information about short-timescale events like the brief outages described in Section~\ref{sec:minieclipses}.} data in a custom format covering 1120 to 1220~MHz, 1320 to 1420~MHz, and 1520 to 1620~MHz.  When possible, we validated observations of J1023 by accompanying them with observations of PSR~J1022+1001, a nearby pulsar with a 10-ms period that rises before and sets after J1023\footnote{J1023 is at the extreme southern limit of the declination range accessible from Arecibo, so it can be observed for no more than about an hour per day.}. For the ASP data, we also took calibration observations with a gated noise diode. Our program consisted of an observation approximately every three weeks, lasting about an hour. This program stopped in early 2010, when the telescope was shut down for repairs, and did not resume thereafter.

Observations with the Westerbork Synthesis Radio Telescope (WSRT) in the Netherlands used the telescope's tied-array mode, which combines (up to) all 14 dishes of the array in phase to provide the equivalent collecting area of a 94-m single dish.  The observations were acquired in one of three bands centered at 150, 350, and 1380~MHz, with respective total recorded bandwidths of 20, 80, and 160~MHz.  The WSRT observations were typically half an hour to a few hours in length.  Baseband data were recorded and subsequently coherently dedispersed and folded offline using the PuMaII pulsar backend \citep{kss08}.

Observations at the Jodrell Bank Observatory in the United Kingdom used a dual-polarization cryogenic receiver on the 76-m Lovell telescope, having a system equivalent flux density of 25~Jy on cold sky.  Data in the range 1350~MHz to 1700~MHz were processed by a digital filterbank for a bandwidth of 112~MHz producing 0.5-MHz channels up until August 2009; subsequently a bandwidth of about 300~MHz with 0.25-MHz~channels was produced. The data were incoherently dedispersed and folded online. Most observations were half an hour long but a few were longer, and in particular one long observation on MJD 54906 was simultaneous with a long WSRT 350-MHz observation.

Observations with the 110-m Robert C. Byrd telescope at Green Bank were carried out with the GUPPI pulsar instrument \citep{rdf+09} in two different modes. For our long observations, we recorded 1100 to 1900~MHz in incoherent dedispersion mode with online folding, producing files in PSRFITS format. Each observation began and ended with a calibration scan using a gated noise diode, though three of the four calibration scans proved unusable. In this configuration we carried out two-full-orbit observations on 2009 May 10 and 2009 May 14. We also carried out a number of short-term low-frequency observations, recording data covering 300 to 400~MHz in incoherently dedispersed search mode and producing data in PSRFITS format \citep{hsm04}.

\section{Long-term timing solutions}
\label{sec:timing}
A key aspect of the current work was maintaining a long-term timing solution for J1023 \citep[for example in support of][]{das+12}. In light of the complexities we will discuss in Section~\ref{sec:radiophenomenology} and below, frequent observations were essential. Early on, the regular 1400-MHz monitoring with the Arecibo telescope provided the key timing observations, while later the Jodrell Bank 1400-MHz observations, at some times almost daily, provided essential timing information. Even with this large collection of observations, obtaining an adequate model for the rotation of J1023 was difficult.

Many observations were obtained in online folding mode, based on our then-best estimate of the pulsar's ephemeris. Others were taken in search mode (power measurements in each frequency channel every few tens of microseconds). We folded the search-mode data with preliminary versions of our long-term ephemeris, then refolded when the ephemeris was improved significantly. The result of this latter process was a folded pulsar observation consisting of one profile per subintegration ($30\;\text{s}$ to several minutes) per frequency channel. 

\begin{figure}
    \plotone{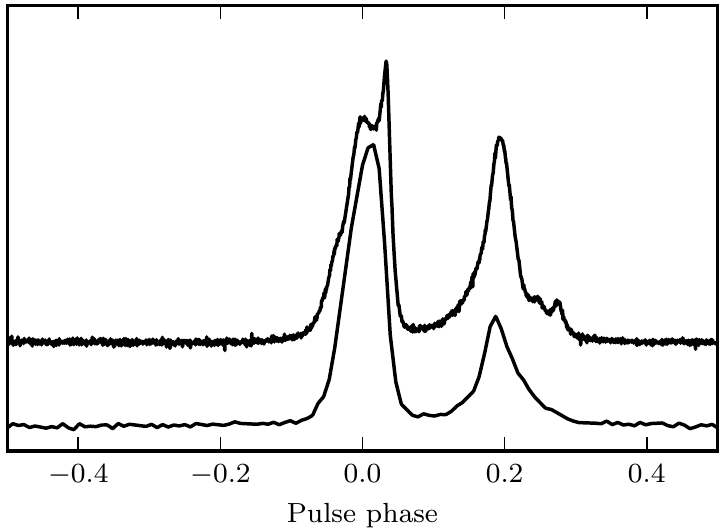}
    \caption{\label{fig:stdprof} Radio pulse profiles of J1023. Top: 1400-MHz profile obtained with coherent dedispersion using the ASP at Arecibo. Bottom: 350-MHz profile obtained with coherent dedispersion using the WSRT. The vertical scale is flux density, scaled arbitrarily for each profile. The profiles are arranged horizontally to approximately align both major peaks.}
\end{figure}
For each combination of instrument and frequency band, we selected an observation in which the folded pulse had a good signal-to-noise ratio and negligible phase drift or DM variation. We then averaged these observations down to a single polarization, a single frequency channel, and a single time integration to produce a high signal-to-noise ratio template (like those in Figure~\ref{fig:stdprof}). We then used this template and either the PSRCHIVE \citep{hsm04} tool \texttt{pat} or the PRESTO \citep{rans01} tool \texttt{get\_toas.py} to produce pulse time-of-arrival measurements (TOAs) by cross-correlation. For some observations we produced two sets of TOAs, one set based on averaging across the whole observing band and the second based on averaging across a modest number of subbands. The former process produced TOAs with lower uncertainties and in a more modest number, while the latter process permitted testing for DM variations (see Section~\ref{sec:excessdm}) but produced too many TOAs for use in a global timing solution.

When observing J1023, a number of phenomena may corrupt individual TOAs. Most obvious are eclipses and flux variations, which may make pulse phase measurements impossible. More subtle are DM variations, which introduce additional frequency-dependent delays. These variations, discussed in Section~\ref{sec:excessdm}, can contribute substantially to the systematic errors in TOA measurements. In the 350-MHz data, these effects can be substantial at all orbital phases. Fortunately, in the 1400-MHz data these effects are small except at orbital phases close to inferior conjunction. We used only the 1400-MHz data for timing purposes, and we handled these various effects by rejecting all TOAs with phases too near inferior conjunction or with measured uncertainties larger than $10\;\mu\text{s}$. We also removed certain TOAs that showed evidence of being severely affected by excess DM (as described in Section~\ref{sec:excessdm}).
\begin{figure*}
    \plotone{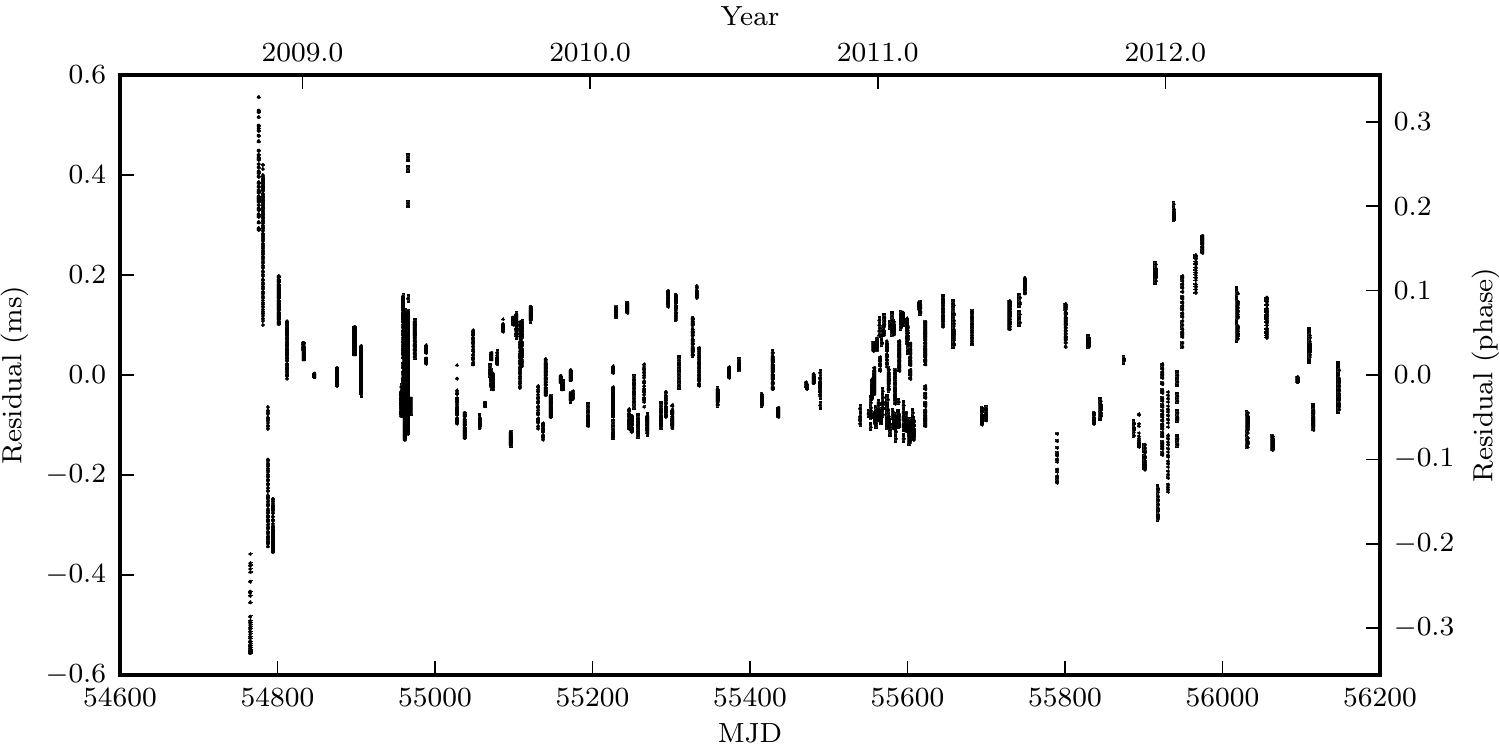}
    \caption{\label{fig:longtermresidsbat} Residuals as a function of time for the simple long-term solution in Table~\ref{tab:longterm}. Uncertainties are marked with vertical bars, which are often smaller than the circle used to mark the value. Because we were able to extract many TOAs from each observing session, there are generally many data points at nearly the same horizontal position. Since the orbital phases of these points differ, the orbital phase dependence of the residuals seen in Figure~\ref{fig:longtermresids} cause them to appear as vertical bars here.}
\end{figure*}
\begin{figure*}
    \plotone{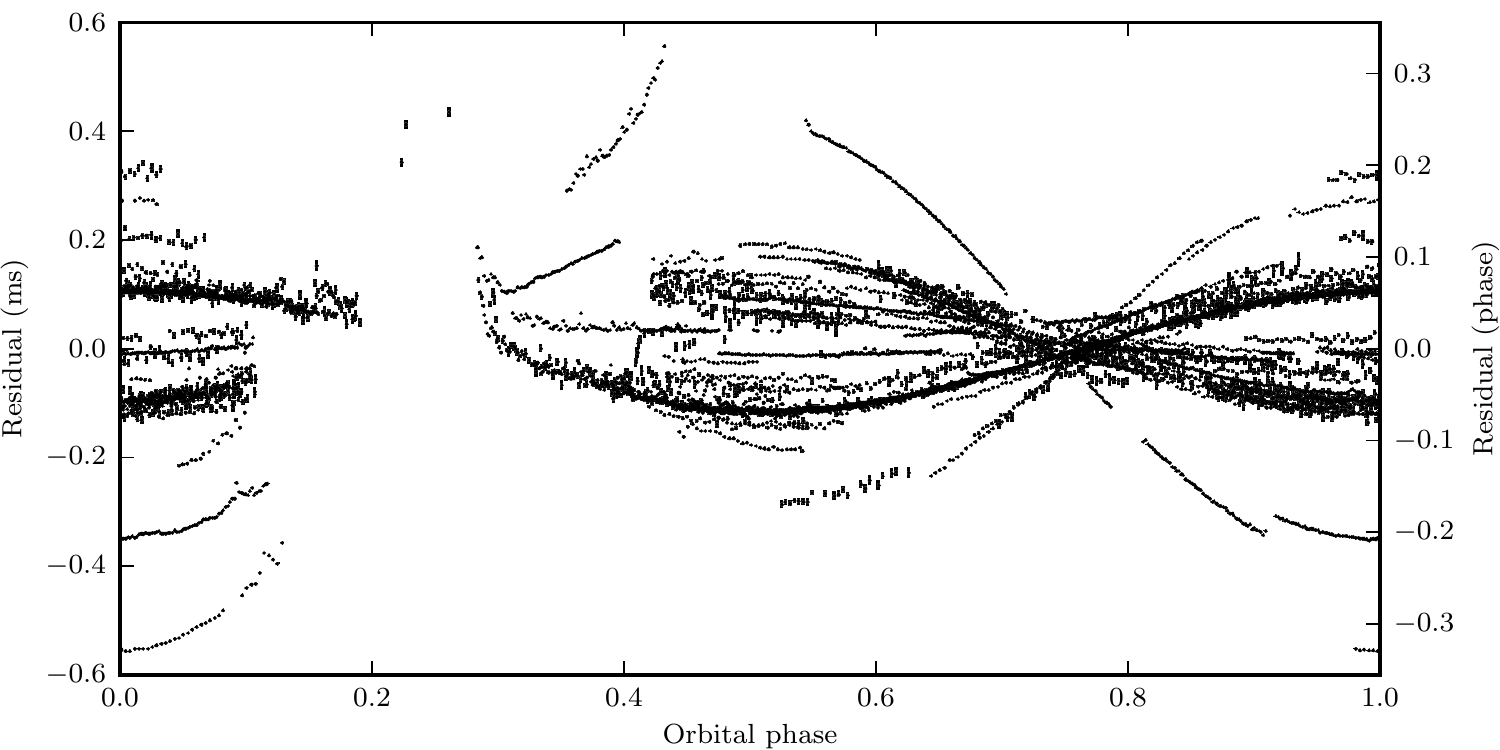}
    \caption{\label{fig:longtermresids} Residuals as a function of orbital phase for the simple long-term solution in Table~\ref{tab:longterm}. Vertical bars indicate uncertainties. In some cases they are smaller than the circle used to mark the value.}
\end{figure*}

Given a collection of good TOAs, the task of producing an ephemeris consists of two parts, in our case carried out using the pulsar timing software \texttt{tempo2} \citep{hem06}. The first is to determine the exact number of turns between each pair of TOAs, while the second adjusts the ephemeris to model the rotation more closely. The first process is called ``phase connection'' and is normally a manual, iterative process of extending a partial ephemeris to cover new observations. Errors in the predicted number of turns generally reveal themselves by producing residuals with errors greater than a few tenths in phase. With the kind of dense sampling possible with J1023, errors in the turn count may also be visible as large trends within single observations. We carried out the process of phase connection on our observations of J1023, producing the ephemeris given in Table~\ref{tab:longterm}. In building this ephemeris, we fixed J1023's position, proper motion, and parallax at the values obtained with VLBI \citep{das+12}. We also fixed the DM at $14.3308\;\text{pc}\;\text{cm}^{-3}$, a value obtained from fitting a WSRT 350-MHz observation that is far from eclipse and that has good signal-to-noise (see Section~\ref{sec:excessdm} for a discussion of the DM variations we observe). The DM selected is relatively unimportant in the long-term fitting, since all the TOAs from a given telescope/backend/band combination are at the same frequency, and we fit for arbitrary offsets between telescope/backend/band combinations.

The ephemeris in Table~\ref{tab:longterm} produced the phase residuals shown in Figures~\ref{fig:longtermresidsbat} and~\ref{fig:longtermresids}. While these residuals are in some cases as large as $0.2$ turns, their smooth orbital dependence makes it clear that the ephemeris does correctly phase-connect all our observations --- that is, we can unambiguously account for the exact number of pulsar rotations between each observing epoch. While the orbital dependence of the residuals indicates unmodelled orbital variations, the scale of these variations is rather small when it comes to orbital phase prediction. As a compromise between model complexity and residual size, this ephemeris uses a very simple orbital model, allowing the orbital period to vary linearly over the span of our observation. We believe this linear variation is not physically significant and does not represent a longer-term trend, instead being the cumulative effect of the random orbital variations described in Section~\ref{sec:timingvariations}. Nevertheless this ephemeris is appropriate for orbital phase predictions and for online-folding observations of J1023 that are not too far in the future.

In the interest of better understanding the orbital variations and other phenomenology observed in J1023, it is valuable to use \texttt{tempo2} to fit many shorter-term observational spans. However, such automated use of \texttt{tempo2} risks introducing a spurious phase turn, rendering useless the results of the fitting procedure. In order to prevent this, we implemented a modification to \texttt{tempo2}. By default, when \texttt{tempo2} computes a residual for a pulse time-of-arrival measurement, it simply chooses the turn number that produces the smallest residual, that is, it attempts to make the residual as close to zero as possible. If the proposed solution is too far from the true solution, so that the residual should be larger than $0.5$ in phase, this results in an extra phase turn.
To mitigate this problem we wrote code to allow us to control the turn number directly.

The \texttt{tempo2} output plugin \texttt{general2} allows the output of turn numbers for each TOA. Although \texttt{tempo2} includes a mode, activated with \texttt{TRACK -2}, which allows turn numbers to be specified as part of the TOA data, this mode suffers from a bug causing turn numbers to wrap around after approximately two billion turns (about 1.4 months for J1023). We therefore implemented a fixed version as \texttt{TRACK -3}, and we have submitted a bug report to the \texttt{tempo2} mailing list. With this mode we were able to use the phase-connected solution from Table~\ref{tab:longterm} and the \texttt{general2} plugin to annotate our TOAs with turn numbers. We were then able to activate \texttt{TRACK -3} and be certain that no automated fit introduced spurious phase turns. We point out that this allows, for example, easy changing from one timing model to a differently parametrized timing model; wrong parameters will not result in phase wraps and can therefore be corrected by \texttt{tempo2}'s fitting procedure.

This mode for \texttt{tempo2} was useful at several stages in our analysis. During initial phase connection, it was useful for stitching together solutions each covering part of the data set: if the differences between turn numbers agreed for those TOAs for which the solutions overlapped, and the overlap was sufficient, then the solutions were compatible, and we could merge the lists of turn numbers to annotate all the TOAs in either data set. Then a timing solution could be fit to the data without the complication of phase wrapping. Second, once we had a phase-connected timing solution, we could ``bake in'' the turn numbers so that the automated short-term fitting we describe in Section~\ref{sec:timingvariations} and Section~\ref{sec:gamma} could not introduce a phase wrap and therefore converge on an incorrect solution.

\section{Radio phenomenology}
\label{sec:radiophenomenology}

\begin{figure*}
    \plotone{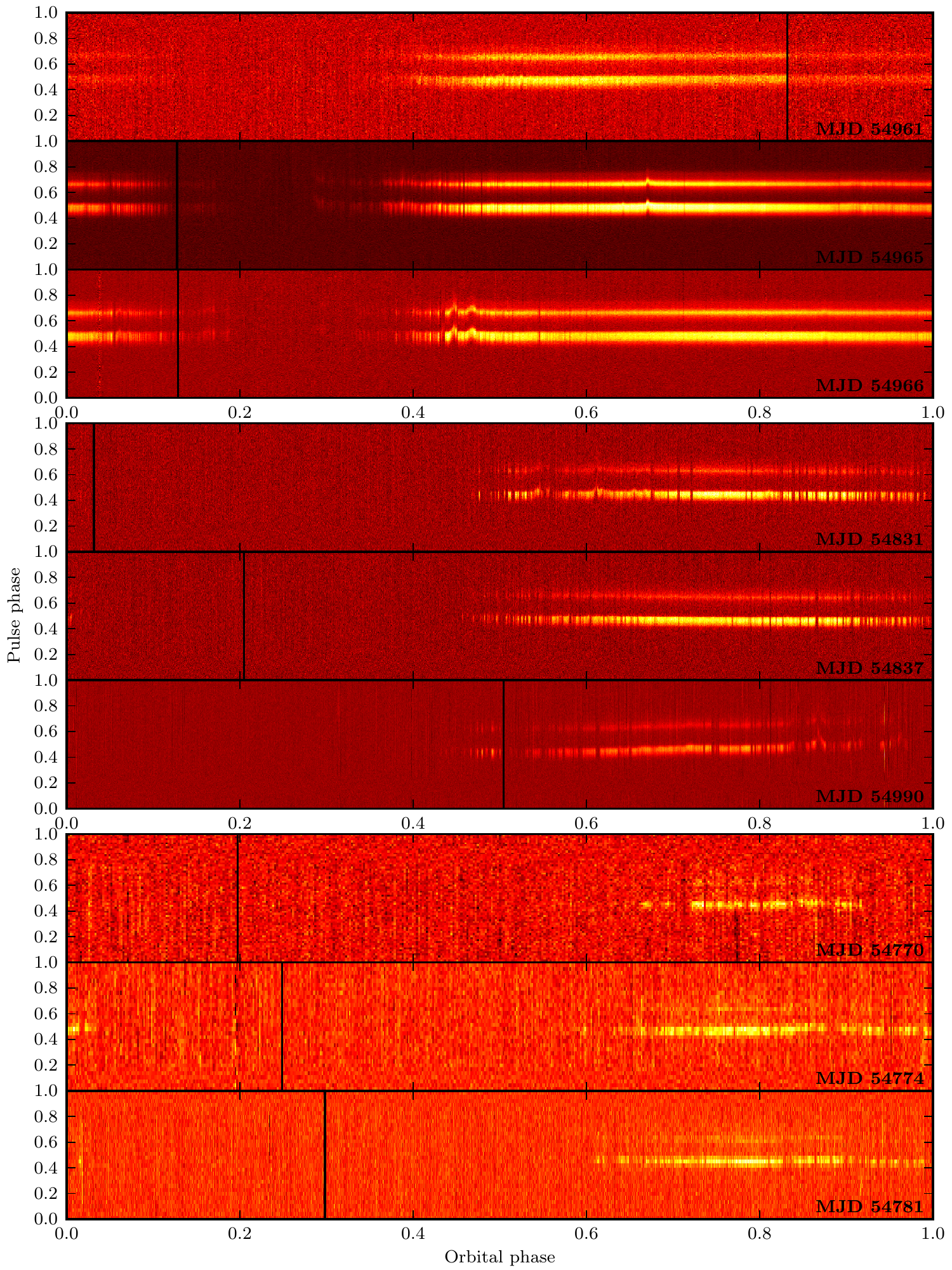}
    \caption{\label{fig:radiooverview}Several full-orbit observations of J1023. Color indicates flux density as a function of pulse and orbital phase. The top group of three panels shows 1400-MHz observations with the GBT, the middle shows 350-MHz observations with the WSRT, and the bottom shows 150-MHz observations with the WSRT. Within each panel, the observation begins with the black vertical line, wraps around, and ends with the black line. The date of the observation is indicated within each panel; the second and third 1400-MHz panels are in fact taken from a single scan, so that the third panel takes up where the second leaves off. For each of the observations below 1400-MHz, we fit for the DM that gave the best signal-to-noise; values varied by $\sim 10^{-3}\;\text{pc}\;\text{cm}^{-3}$. Differences between panels at the same frequency show the orbit-to-orbit variability of J1023; differences between observations at different frequencies show the same combined with the frequency dependence of the various effects described in Section~\ref{sec:radiophenomenology}. For a truly simultaneous dual-frequency comparison see Figure~\ref{fig:eclipse}.}
\end{figure*}

J1023 exhibits a number of phenomena which give some insight into the nature of this complex binary system, several of which are visible in Figure~\ref{fig:radiooverview}. The most obvious is the main eclipse, which occurs near inferior conjunction (orbital phase 0.25), when the companion passes nearest to our line of sight to the pulsar. This eclipse is accompanied by excess DM at ingress and egress. Excess DM is also seen at other orbital phases, varying from orbit to orbit, as is seen in PSR~J1748-2446A \citep[Terzan 5 A;][]{nttf90}. In 150- and 350-MHz observations, we observe brief (seconds to minutes) periods during which the pulsar signal disappears. We also observe the effects of interstellar scintillation on the signal from J1023, which can complicate observing by introducing large intensity variations at 1400-MHz. Finally, the orbital parameters of J1023 are variable, severely complicating long-term timing. 

\subsection{Main eclipse}
\label{sec:maineclipse}
\begin{figure}
    \plotone{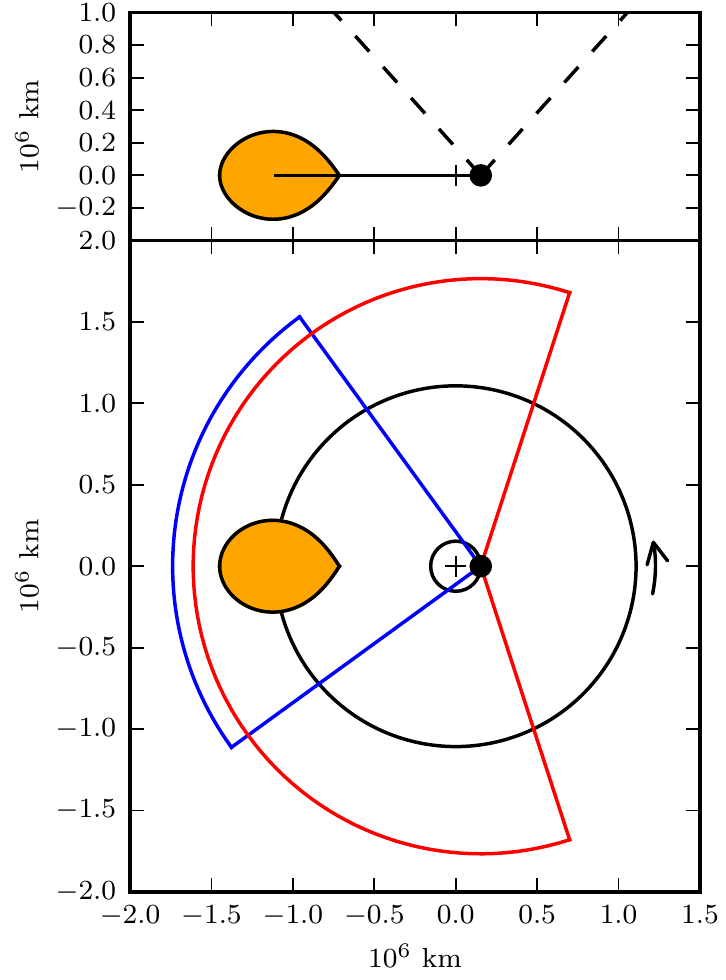}
    \caption{\label{fig:roche}Geometry of the J1023 system and its eclipse. Orange teardrop shows the shape and size of the companion (assumed to fill its Roche lobe), while black dot indicates the pulsar (not to scale). Top: Edge-on view of the system; dashed lines indicate the line-of-sight to Earth at orbital phases 0.25 and 0.75. Bottom: Face-on view of the system. Arcs indicate the main eclipse phase ranges from the dual-frequency observation, for a system rotating counter-clockwise. Red is for the 350-MHz eclipse while blue is for the 1400-MHz eclipse.}
\end{figure}
The most obvious feature of J1023's radio emission is the main eclipse. The system's viewing geometry is well known: based on the mass measurement of $1.71\pm 0.16 M_\Sun$ in \citet{das+12} and the Keplerian mass function of the system, we can infer that the inclination angle is $42\pm 2\degree$. With this geometry, pictured in Figure~\ref{fig:roche}, the companion's Roche lobe does not cross the line of sight to the pulsar. In fact, Figure~1 of \citet{asr+09} shows a 3000-MHz full-orbit observation of J1023 in which the main eclipse is absent: approximately constant emission is visible throughout the orbit. The eclipse is therefore presumably due to relatively tenuous ionized material outside the companion's Roche lobe. On the other hand, at 1400~MHz the phase-averaged flux during eclipse is lower by a factor of several (in fact undetectable) compared to the flux outside eclipse (Deller 2012, personal communication). \citet{asr+09} also showed that the eclipse duration becomes longer at lower frequency.

\begin{figure*} %
    \plotone{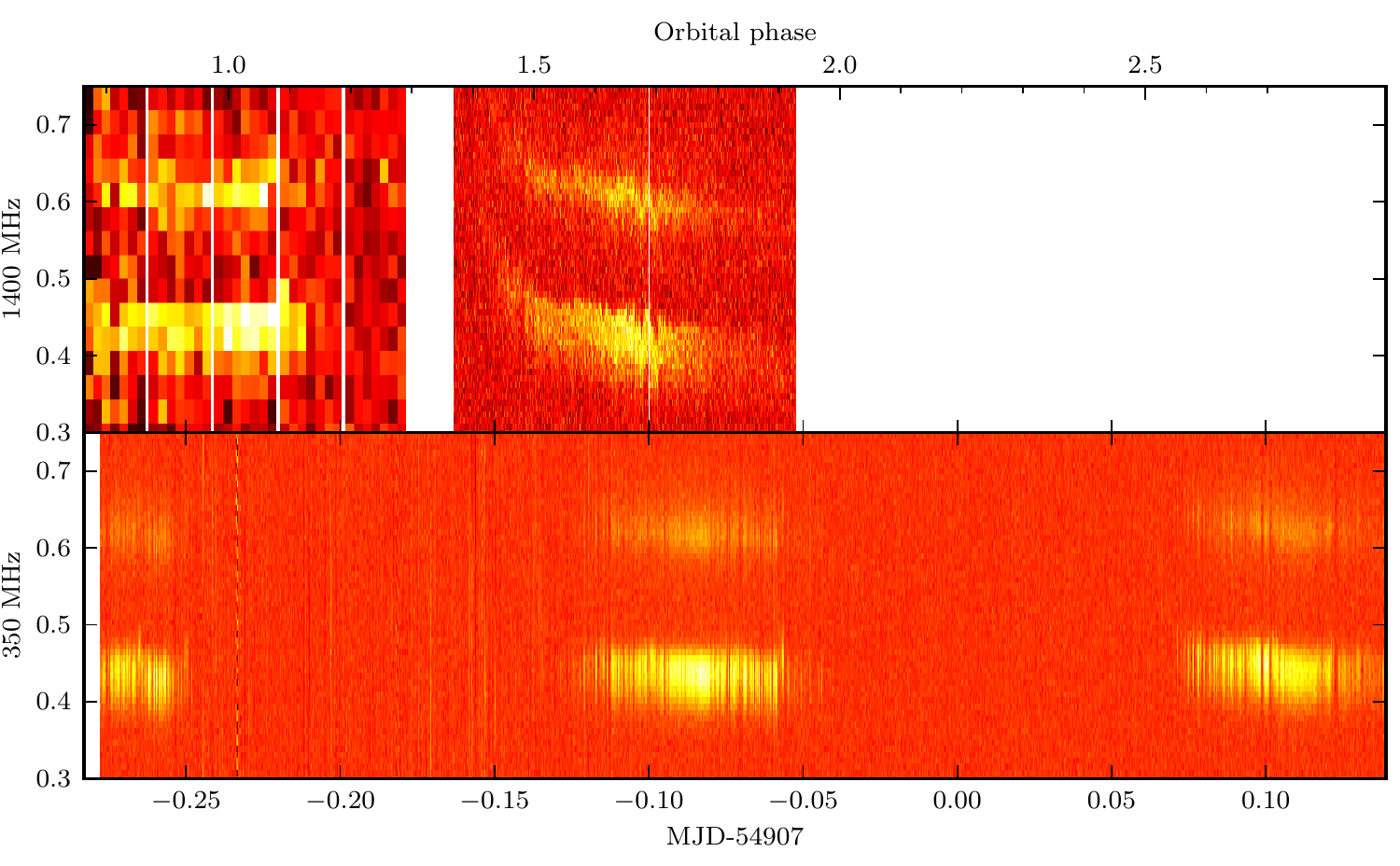}
    \caption{\label{fig:eclipse}Simultaneous dual-frequency observation of J1023. Color indicates intensity as a function of time and pulse phase. The top panel is observed with Jodrell Bank at 1400~MHz, and the bottom panel is observed with the WSRT at 350~MHz. Note that scintillation resulted in a drastically reduced signal-to-noise during the first part of the Jodrell Bank observation, so we have reduced the time and phase resolution.}
\end{figure*}
A dual-frequency observation we acquired allows us to examine this effect more closely. Figure~\ref{fig:eclipse} shows both 350-MHz and 1400-MHz observations taken simultaneously with the WSRT and the Lovell telescope respectively. At 1400~MHz the eclipse length is about $0.25$ in orbital phase, while at 350~MHz it is closer to $0.6$ in orbital phase. It can also be seen that the phase of eclipse ingress varies more than the phase of eclipse egress as observations move to lower frequency, confirming what is seen in the non-simultaneous observations shown in Figure~\ref{fig:radiooverview} and in Figure~1 of \citet{asr+09}.

\subsection{Excess DM}
\label{sec:excessdm}
\begin{figure}
    \plotone{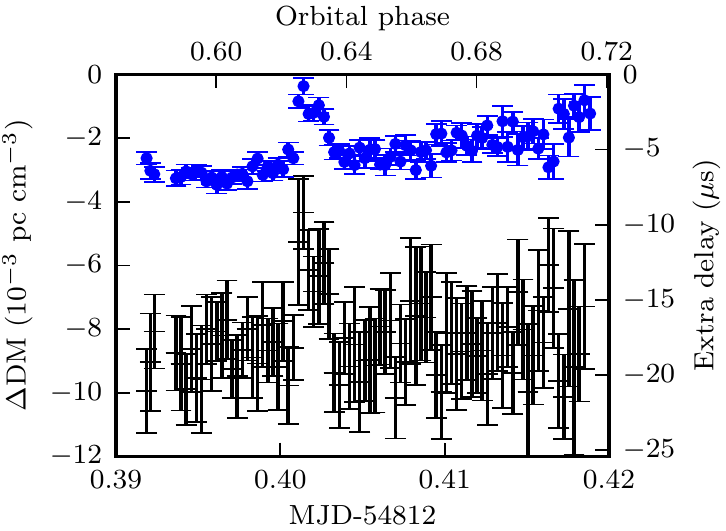}
    \caption{\label{fig:dmexcess}Short-term DM fits to an Arecibo 1400-MHz observation of J1023 on MJD 54812. The vertical error bars show the uncertainty returned by \texttt{tempo2}. The top sequence of points with blue circular markers are simple timing residuals from single-frequency observations assuming a single DM; their amplitude can be read off the scale to the right. The lower sequence of points with black crosses for markers are fits for dispersion measure within single subintegrations; their scale can be read off the left side of the plot. The two scales have been arranged so that the excess dispersion measure on the left corresponds to the excess delay on the right for a 1400-MHz observation.}
\end{figure}
\begin{figure}
    \plotone{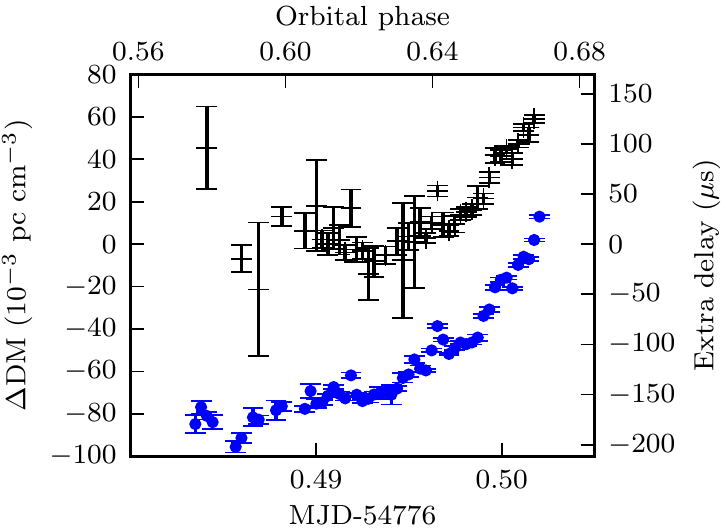}
    \caption{\label{fig:dmrise}Short-term DM fits to an Arecibo 1400-MHz observation of J1023 on MJD 54776. Scales and traces are defined as in Figure~\ref{fig:dmexcess}, but note that because this observation was very near the beginning of our program and cannot be used for timing because of the strong excess DM, there may be a trend superimposed on the timing residuals. Note also that this observation occurs after the main eclipse but shows a very substantial increase in DM.}
\end{figure}
A second feature of J1023's radio emission is excess DM. This is visible in Figure~\ref{fig:radiooverview} and in residual plots as excess delays. Wide-band observations like those taken as part of our monitoring program at Arecibo allow us to confirm that these delays are in fact due to excess DM by computing TOAs from several subbands in the same subintegration, then by fitting for the DM within that single subintegration. The excess DM we observe takes several forms. First, we consistently observe excess DM surrounding the main eclipse. Specifically, we see the DM increase rapidly up to the moment the signal disappears, then fall rapidly down to the normal value after the signal reappears. The amount of excess DM is difficult to estimate, since the points at which we lose and reacquire signal depend on signal-to-noise, which varies substantially due to scintillation. Nevertheless it is clear that the amount of excess DM at a phase near the main eclipse also varies from orbit to orbit. Typical excess DM at loss of signal due to eclipse ingress is roughly $0.01\;\text{pc}\;\text{cm}^{-3}$, while at signal reacquisition due to eclipse egress it is much larger, usually around $0.15\;\text{pc}\;\text{cm}^{-3}$.  Second, in our hour-long monitoring observations, the average DM varies by a few times $10^{-3}\;\text{pc}\;\text{cm}^{-3}$ from what we have adopted as the baseline DM for the system. Finally, in several cases, one of which is pictured in Figure~\ref{fig:dmexcess}, we see short-term (varying lengths but typically $\sim 300\;\text{s}$) excesses of DM (by varying amounts but typically $\sim4\times 10^{-3}\;\text{pc}\;\text{cm}^{-3}$). Importantly, these DM excesses occur far from the main eclipse, and at different orbital phases, as do the signal disappearances discussed in Section~\ref{sec:minieclipses}. We also see one example of a substantially larger and longer-lasting increase in DM, shown in Figure~\ref{fig:dmrise}; on MJD 54776, we observed the DM rise steadily by $\sim 5\times 10^{-2}\;\text{pc}\;\text{cm}^{-3}$ over the course of twenty minutes; our observation ended while the rise continued. This occurred around orbital phase 0.5, after the eclipse.

It should be noted that the delays introduced by this excess DM at 1400~MHz are on the order of tens of microseconds, while the residuals pictured in Figures~\ref{fig:longtermresidsbat} and~\ref{fig:longtermresids} are on the order hundreds of microseconds, so excess DM is not sufficient to explain them. On the other hand, we should expect the effect of DM variations to be some sixteen times greater for 350-MHz observations, producing delays on the order of hundreds of microseconds and phase shifts on the order of tenths of a turn. For this reason, it would be difficult to maintain a good timing solution for J1023 using only 350-MHz data, although given the high signal-to-noise ratio, it is possible to measure the DM and possibly correct for the delays within each observation. In fact, in order to obtain a good signal-to-noise ratio at 350~MHz and below, it is necessary to measure a separate DM for use in each observation.

\subsection{Short-term signal disappearances}
\label{sec:minieclipses}
\begin{figure}
    \plotone{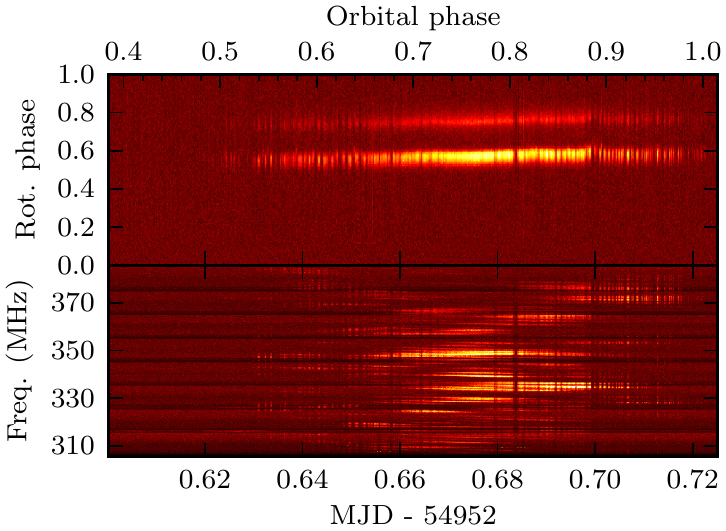}
    \caption{\label{fig:dynamic}Top: Brightness as a function of time and pulse phase for a WSRT 350-MHz observation taken on MJD 54952. Note both short-term signal disappearances and episodes of delayed pulses, presumably due to excess DM. Bottom: Dynamic spectrum of the same observation. Note that the disappearances of signal are broadband and clearly distinct from the scintillation. This observation is assembled from eight subbands, so certain band-edge artifacts are visible, as is some radio-frequency interference.}
\end{figure}
At 150 and 350~MHz, we observe brief periods during which the pulsar signal disappears entirely. In some of our search-mode observations it is possible to see disappearances as short as a few seconds, but in most of our data (folded in 10-s subintegrations) we see disappearances varying in length from one subintegration to as long as minutes. These disappearances occur at all orbital phases (apart from during the main eclipse, when there is no signal to disappear), though they may be more frequent nearer to eclipse. In a few cases the signal disappearances begin or end with excess delays (presumably due to excess DM). The dynamic spectrum (see Figure~\ref{fig:dynamic}) makes it clear that the disappearances span the entire bandpass ($\sim 80$~MHz). As visible in Figure~\ref{fig:radiooverview}, these signal disappearances also occur at 1400~MHz and 150~MHz. That said, in our 350-MHz/1400-MHz dual-frequency observation, disappearances that occur at the lower frequency are not accompanied by disappearances at 1400~MHz; we are not able to test the converse because no short disappearances are present in our simultaneous observation. 

\subsection{Interstellar Scintillation}
\label{sec:scintillation}
Interstellar scintillation results when the turbulent interstellar medium modulates the radio flux from a pulsar. The variation appears as a collection of \emph{scintles} with a characteristic duration and bandwidth and with varying intensity. A few of these scintles, detected in our 350-MHz data, are visible in Figure~\ref{fig:dynamic}. An autocorrelation analysis \citep{lk04} shows a bandwidth $\Delta f_{DISS} $ of $1.4\;\text{MHz}$ and a timescale $\Delta t_{DISS}$ of $1100\;\text{s}$. The errors on these quantities are dominated by the small number of bright scintles. Following \citet{cord86}, we estimate the number of scintles in the observation we used as the total number of possible scintles times a small filling factor to account for the dominance of exceptionally bright scintles in our autocorrelation analysis: $N_s = 10^{-2}(T/\Delta t_{DISS})(B/\delta f_{DISS}).$ In this case we are using a $T=7000\;\text{s}$ observation segment (counting only non-eclipsed time) with $B=100\;\text{MHz}$, so we estimate $N_s = 4.5$; we should therefore expect fractional errors on scintillation parameters of $N_s^{-1/2} = 0.5$. Nevertheless, \citet{lk04}, citing \citet{cr98}, give the following formula for inferring the pulsar's proper motion from scintillation parameters:
\begin{align*}
    V_{\text{ISS}} = & 2.53\times 10^4~\text{km}\;\text{s}^{-1}
    \left( \frac{d}{1\text{kpc}} \right)^{1/2} \times \\
    & \left( \frac{\Delta f_{\text{DISS}}}{1\text{MHz}} \right)^{1/2}
    \left( \frac{ f}{1\text{GHz}} \right)^{-1}
    \left( \frac{\Delta t_{\text{DISS}}}{1\text{s}} \right)^{-1}.
\end{align*}
Applying this to the known distance \citep[$1.37\;\text{kpc}$;][]{das+12} and a frequency of $350\;\text{MHz}$ gives a velocity of $90\;\text{km}\;\text{s}^{-1}$. Assuming the fractional error of $50\%$ computed above makes this roughly consistent with the observed proper motion of $130\;\text{km}\;\text{s}^{-1}$ \citep{das+12}, particularly as during this partial orbit the speed at which the sight line sweeps through the interstellar medium is also affected by the pulsar orbital motion, which varies with orbital phase from $40$ to $60\;\text{km}\;\text{s}^{-1}$ and is in an unknown direction.  At $1400\;\text{MHz}$ the scintles are too large to obtain good statistics, but visual inspection suggests a scintillation bandwidth of $\sim 50\;\text{MHz}$ and timescale of $\sim 1\;\text{hour}$; this implies a velocity of $\sim 40\;\text{km}\;\text{s}^{-1}$, with an even larger fractional uncertainty given the presence of only a handful of scintles within an observation. In practical terms, interstellar scintillation introduces a wide variation in observed radio flux at $1400\;\text{MHz}$, particularly for narrow-band observations, for example those described in \citet{das+12}, but also visible in the broadband observations shown in Figure~\ref{fig:radiooverview}.

\subsection{Timing Variations}
\label{sec:timingvariations}
\begin{figure}
    \plotone{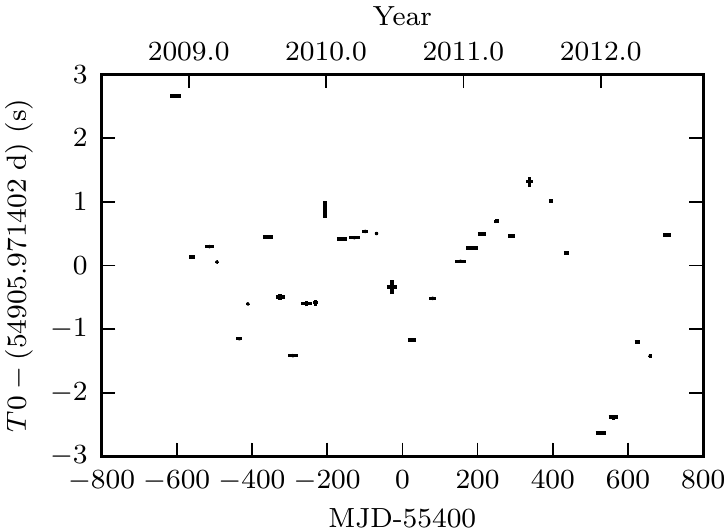}
    \caption{\label{fig:orbvariations}O-C (observed minus calculated) diagram of the time of orbital phase 0. A version of the long-term ephemeris was fit to each approximately 30-day section of data, allowing only orbital phase (and pulse phase) to vary, and locking the number of pulsar rotations to that computed from the original long-term ephemeris. Vertical bars indicate uncertainties reported by \texttt{tempo2} except where they are smaller than the circle used to indicate the value; horizontal bars indicate the range of data used in the fit. Since the original ephemeris was obtained by fitting an orbital period and its derivative to the whole data span, we have effectively removed the best-fit quadratic from these data points. Residual scatter appears to come from intrinsic orbital period variations.}
\end{figure}
In addition to pulse delays due to excess DM, timing of J1023 is made difficult by its erratic orbital behavior. Figure~\ref{fig:longtermresidsbat} shows the poor quality of the fit obtainable by a simple model of the orbit including only a secular change in orbital period. Figure~\ref{fig:longtermresids} also shows that the residuals have a substantial dependence on orbital phase, suggesting that varying the orbital parameters might improve the fit. While \texttt{tempo} offers a binary model allowing many orbital period derivatives, porting it to \texttt{tempo2} and fitting numerous orbital period derivatives failed to produce satisfactorily small residuals. Instead, we chose to fit short-term piecewise solutions, in which we allowed orbital phase to vary. When using non-overlapping 30-day intervals, we obtained root-mean-squared residuals typically less than 1\% of a turn. We also experimented with fitting binary period and/or amplitude within each segment, which reduced the residuals by a small additional amount. The fitted orbital parameters showed no discernable structure, either by eye or using a Lomb-Scargle periodogram, even when using intervals as short as 14 days. The scale of the orbital phase variations required for our 30-day fits is shown in Figure~\ref{fig:orbvariations}; they are of order $1\;\text{s}$.

In summary, the orbital variations we observe are substantial but difficult to
quantify, let alone understand. In Section~\ref{sec:discorbital} we will discuss possible explanations for these orbital variations.

\section{\texorpdfstring{$\gamma$}{Gamma}-ray behavior}
\label{sec:gamma}

The \emph{Fermi} Large Area Telescope (LAT) is a sky-scanning $\gamma$-ray telescope sensitive to photons of ${\sim}100$ MeV to $300$ GeV, and it has proved a powerful tool for studying pulsars. It is described in \citet{aaa+09}, and its on-orbit performance is detailed in \citet{aaa+12}. \citet{thh+10} discovered a point source in data from the LAT coincident with the position of J1023, and indeed J1023 is associated to a source in the second \emph{Fermi}-LAT source catalog, namely 2FGL~J1023.6+0040, with a 1--100 GeV energy flux of $(5.4\pm 0.9)\times 10^{-12}\;\text{erg}\;\text{cm}^{-2}\;\text{s}^{-1}$ and a power-law index of $(2.5\pm 0.3)$ \citep{naa+12}. \citet{thh+10} also report an energy flux of $(5.5\pm 0.9)\times 10^{-12}\;\text{erg}\;\text{cm}^{-2}\;\text{s}^{-1}$ for $>200\;\text{MeV}$. Either value corresponds to a $\gamma$-ray luminosity of $1.2\times 10^{33}\;\text{erg}\;\text{s}^{-1}$. To better understand this $\gamma$-ray emission, we have obtained and folded the \emph{Fermi} LAT photons apparently coming from J1023. 

\subsection{Data}
We requested pass 7 photon data from the \emph{Fermi} data server for the energy range 100 MeV to 300 GeV and a circular region of radius $15\degree$ centered on J1023's position.  The date range we selected was from 2008 August 4 to 2012 August 18. Our data selection follows the \emph{Fermi} Cicerone\footnote{\url{http://fermi.gsfc.nasa.gov/ssc/data/analysis/documentation/Cicerone/}}. We used photons from the \texttt{P7SOURCE} class; to reduce contamination from the Earth limb we required that the zenith angle be less than $100\degree$. We selected time ranges corresponding to good quality data taken during sky-survey operation by applying the expression \texttt{(DATA\_QUAL==1) \&\& (LAT\_CONFIG==1) \&\& ABS(ROCK\_ANGLE)<52} with \texttt{gtmktime}; we also used \texttt{gtmktime} to select time ranges for which the entire $15\degree$ ROI was less than $100\degree$ from zenith. Using the \emph{Fermi} tool \texttt{gtsrcprob} and a spectral model for all sources within $5\degree$ of the ROI from the 2FGL survey (including the \texttt{P7SOURCE\_V6} instrument response functions, the diffuse model \texttt{gal\_2yearp7v6\_v0.fits} and the isotropic model \texttt{iso\_p7v6source.txt}), we assigned to each photon a probability that it came from J1023 then culled all photons with probability less than $10^{-3}$.  This culling reduced the number of photons needing processing from 622882 to 47032 and the number of expected photons from the source from 623.6 to 521.6. Our Monte Carlo simulations make it clear that the low-probability photons we discarded make a negligible difference to detection significances. 

\subsection{Orbital modulation}
To look for orbital modulation, we used the simple orbital ephemeris given in Table~\ref{tab:longterm} and the \texttt{tempo2} \texttt{fermi} plugin to compute the orbital phase at which each photon was emitted. We used a weighted form of the H test \citep{kerr11} and found a false positive probability of 0.78. The highest degree of orbital modulation consistent with this detection significance depends on the hypothetical pulse shape. As it takes into account multiple harmonics, the H test is least sensitive to a simple sinusoid. To estimate an upper limit on the fraction of the photons that could be pulsed, we ran simulations of sinusoidally modulated signals. We found that if $35\%$ of the source photons were pulsed, there would be at least a $95\%$ chance that we would have obtained a result more significant than what we actually observed.
\subsection{Rotational modulation}
\begin{figure}
    \plotone{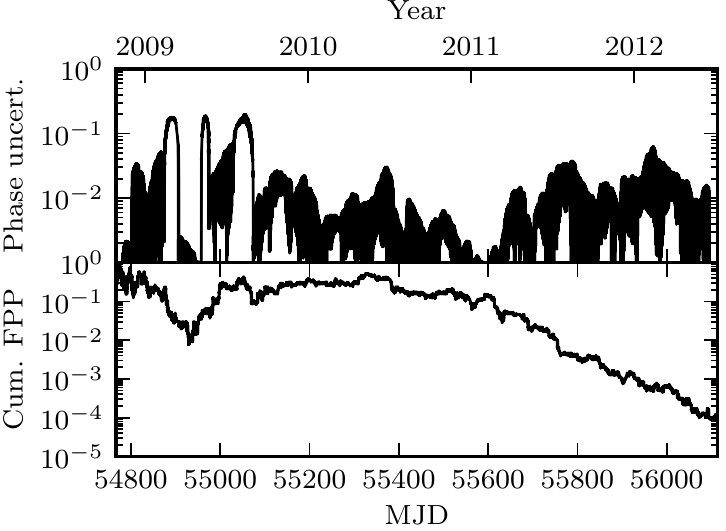}
    \caption{\label{fig:fppscatter}Top panel: Estimated phase uncertainty. We computed and plotted the weighted standard deviation of the phase predictions at the time of each \emph{Fermi} LAT photon. Bottom panel: Cumulative false positive probability for pulse phase. Values are based on the weighted H test applied to all usable \emph{Fermi} LAT photons before a given MJD.}
\end{figure}

To look for rotational modulation of the $\gamma$-rays, we needed a reliable phase prediction for each pulse. To model the complex rotational behavior of J1023, we assembled a collection of overlapping ephemerides. Each ephemeris spanned 90 days, and a new ephemeris was started approximately every 14 days. We used the \texttt{fermi} plug-in for \texttt{tempo2} to assign an arrival phase to each photon for each such ephemeris. For each photon and each ephemeris, we assigned a weight between zero and one based on its proximity to the middle of the interval used to fit the ephemeris. Photons outside the interval were assigned zero weight, and photons whose total weight was zero were discarded, limiting the photons used to the times covered by our radio timing observations, MJDs 54766--56114. We computed a weighted mean phase for each photon, effectively producing a piecewise ephemeris predicting rotational phase. The weighting procedure ensured that the predicted ephemeris was continuous and phase predictions were dominated by piecewise ephemerides using observations surrounding the prediction time. We also computed a weighted standard deviation as an estimate of phase uncertainties, plotted in Figure~\ref{fig:fppscatter}. All such standard deviations are less than about $10\%$ in phase, and most are below $1\%$, indicating that ephemeris uncertainties probably do not smear the pulse profile significantly below approximately the tenth harmonic. This collection of piecewise ephemerides, or the spliced combination, is appropriate for short-term phase prediction throughout the span of time it covers.

\begin{figure}
    \plotone{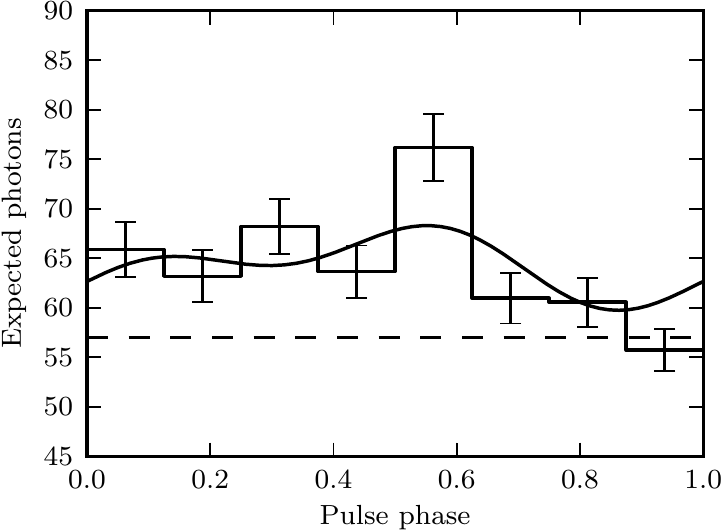}
    \caption{\label{fig:rotational}$\gamma$-ray light curve for J1023 as a function of rotational phase, based on weighted \emph{Fermi} LAT photons. Sinusoidal curve is the best fit using the number of harmonics suggested by the H test. Uncertainties on histogram values are the quadrature sum of all photon probabilities \protect \citep{pga+12}, while background level (dashed line) is the weighted sum of probabilities that the photon is not from the pulsar \protect \citep{gjv+12}.}
\end{figure}

The resulting profile is plotted in Figure~\ref{fig:rotational}, again in the form of a weighted histogram. The H test produces a false positive probability of $9.2\times 10^{-5}$, equivalent to a significance of $3.7\sigma$. The best-fit truncated Fourier series has 2 harmonics and is plotted over the histogram. These significances are single-trial significances, appropriate because we carried out only a single trial, since both the ephemeris and the spectral model were given. We examined the dependence of the significance on energy and time cuts: a very modest improvement in significance (to $4.2\sigma$) is available by discarding all photons below $400~\text{MeV}$, possibly because these are more poorly localized and therefore more contaminated by nearby sources. When we compute a cumulative detection significance (see Figure~\ref{fig:fppscatter}) it wanders for the first year, during which our ephemeris has the largest uncertainty, and increases steadily after that. We estimated the degree of pulse modulation that would lead to such a false positive probability by running Monte Carlo simulations taking into account the uncertainty in which photons come from the source and assuming that photon arrival phases are scattered independently by the phase uncertainty we calculated for each photon. Assuming a sinusoidal pulse profile affecting $90\%$ of the source photons, the observed significance is greater than that observed in about $50\%$ of the simulations. 

\subsection{Summary}
In summary, we detect no orbital modulation, with an upper limit of about $35\%$ on the fraction of photons that are pulsed. We detect weak evidence (statistically $3.7\sigma$) for pulsations at the pulsar's rotational period. Even such a marginal detection, if real, would suggest that most of the photons participate in the modulation.

\section{Discussion}
\label{sec:discussion}

J1023 is at a pivotal moment in its evolution from X-ray binary to MSP. \citet{ccth13} model the evolution of X-ray binaries into redbacks and black widows, finding that the key difference between the two classes is that the redbacks ablate their companions more effectively than the black widows. Key to their simulated evolutionary histories is the idea that low-mass X-ray binaries continue accretion driven by magnetic braking up to the point where the companion becomes fully convective, at which point it loses its magnetic field and accretion stops \citep{rvj83}. For a few highly compact systems, gravitational radiation shrinks the orbit and drives continued mass transfer, but for systems like J1023 that are not so compact, this is the moment where \citet{ccth13} postulate the system turning on as a radio pulsar. In their model, at this point, accretion ceases and is completely replaced by ablation of the companion by the pulsar wind. They describe J1023 as a reasonably typical redback undergoing cyclic episodes of accretion, possibly related to a combination of donor star irradiation \citep{br04}, disc instabilities \citep{dlhc99}, and the radio ejection mechanism \citep{bpd+01}. In particular they view J1023's companion as being non-magnetic. Is a non-magnetic companion compatible with the phenomena we described above? We will try to answer this in the next few sections. 

A second key puzzle is the nature and cause of J1023's active phase. In the 2001 episode, we know that the optical emission grew brighter, bluer, and developed double-peaked emission lines, but X-ray upper limits indicate that the system did not undergo full-fledged accretion \citep{asr+09}. We do not know what caused this active phase or what was happening while it went on.  Based on J1023's behavior in its current, quiescent, state, we will discuss several mechanisms that might lead to such episodic activity.

In fact, while this paper was in internal refereeing, J1023 appears to have entered a new active phase: it disappeared in radio \citep[][Stappers et al. in prep]{sab+13}, it brightened by a factor of five in $\gamma$-rays \citep[][Stappers et al. in prep]{sab+13}, and it shows brighter, variable emission in X-rays and ultraviolet \citep{pah+13}. This is all consistent with the 2000--2001 active phase, offering the chance to study J1023 while active.

\subsection{Basic parameters}
\label{sec:discussbasic}

\citet{asr+09} described the J1023 system based on only a few months of radio observations (and its optical history). Since then, its X-ray and $\gamma$-ray properties have been reported \citep{akb+10,bah+11,thh+10}, and a precise distance measurement has been carried out \citep{das+12}. It seems valuable, then, to summarize the basic system parameters in light of all these data plus our own long-term timing.

Although the orbital parameters vary, as discussed in Section~\ref{sec:timingvariations}, these variations are tiny in absolute terms. Combined with the mass measurement obtained by \citet{das+12}, we can describe the system geometry more or less completely. In particular, the system inclination is $42\pm 2\degree$, and the separation between the companion's center of mass and the pulsar is $1.3\times 10^6\; \text{km}$ ($1.9R_\Sun$). The system geometry is illustrated in Figure~\ref{fig:roche}, along with the ranges of eclipse phase we observed in our dual-frequency observation.

This orbital geometry was inferred from the mass measurement in \citet{das+12}. This mass measurement depended on a key assumption: that the companion filled its Roche lobe. This allowed a measurement of the companion size to be converted into a measurement of the Roche lobe size and therefore the system masses. Modelling of the optical light curves of another very similar redback, PSR~J2215+5135, constrains its companion's Roche lobe filling factor to be $0.99\pm 0.03$ \citep{bkr+13}. While no such fitting process has yet been carried out for J1023, we can make some inferences about the system nonetheless. If the companion underfills its Roche lobe, the masses and system geometry will be different from those assumed elsewhere in this paper. Specifically, the masses will be larger than those reported in \citet{das+12}, and the system will be closer to face-on (the inclination angle $i$ will be less than $42\degree$). That said, it seems unlikely that any pulsar can have a mass larger than $3M_\sun$ \citep{lp04}; this limits the Roche lobe filling factor (companion radius over Roche lobe radius) to be larger than $0.82$, and $i$ to be at least $36\degree$. 

The measured period derivative, $(6.930\pm 0.008)\times 10^{-21}$, must be corrected for the Shklovskii effect \citep{shkl70} and the effect of acceleration due to the Galactic gravitational field. Both these corrections are computed in \citet{das+12}; applying them to our period derivative measurement gives an intrinsic period derivative of $(5.39\pm 0.05)\times 10^{-21}$. Using the standard nominal moment of inertia for pulsar calculations ($10^{45}\;\text{g}\;\text{cm}^2$) we can calculate a spin-down luminosity of $(4.43\pm 0.04)\times 10^{34}\;\text{erg}\;\text{s}^{-1}$. The magnetic field we infer, using the standard formula $B=3.2\times 10^{19}\;\text{G}\;\sqrt{(P/1\;\text{s})\dot P}$, is $9.7\times 10^7\;\text{G}$, and the characteristic age $\tau=P/2\dot P$ is $(4.96\pm 0.05)\times 10^9\;\text{y}$.

\subsection{Energetics}
\label{sec:energy}

Given the distance measurement of \citet{das+12}, we are able to estimate an energy budget for the pulsar. The total spin-down power is $4.43\times 10^{34}\;\text{erg}\;\text{s}^{-1}$ ($11.5 L_\sun$). For comparison, the companion's bolometric luminosity is approximately $6\times 10^{32}\;\text{erg}\;\text{s}^{-1}$ ($0.2 L_\sun$), based on the temperature from \citet{ta05} and the radius of the Roche lobe. 

\citet{ta05} estimated that the amount of irradiation received by the companion was consistent with the compact object having an isotropic luminosity of $2 L_\sun$ ($8\times 10^{33}\;\text{erg}\;\text{s}^{-1}$); we therefore infer an irradiation efficiency of $20\%$. That said, the companion only intercepts about $1.3\%$ of any isotropic emission, so that the actual heating of the companion is closer to $10^{32}\;\text{erg}\;\text{s}^{-1}$ ($0.03L_\Sun$; about $20\%$ of the companion's bolometric luminosity). The average X-ray flux reported in \citet{bah+11} (choosing a power-law plus hydrogen atmosphere spectral model from among their possible assumptions) corresponds to an X-ray luminosity of $9.3\times 10^{31}\;\text{erg}\;\text{s}^{-1}$ ($0.024 L_\sun$), and an efficiency of $0.21\%$.  The $\gamma$-ray luminosity deduced from the $\gamma$-ray flux reported in the 2FGL catalog \citep{naa+12}, $1.2\times 10^{33}\;\text{erg}\;\text{s}^{-1}$ ($0.3 L_\sun$), is  $3\%$ of the nominal spin-down power. 

Comparing these efficiencies to those of the general population of rotation-powered pulsars reported in \citet{vby11}, we find the X-ray efficiency of J1023 is three times the typical value, but within the scatter (though their data includes no MSPs; they report on point-source plus nebular emission, and since \citet{akb+10} and \citet{bah+11} detect no extended emission from J1023, if there is nebular emission it is included in the point source values given above). The $\gamma$-ray efficiency of J1023 is low but not unreasonably so compared to those reported for MSPs in \citet{aaa+13}, including MSPs with higher and lower spin-down powers; J1023's $\gamma$-ray efficiency similarly fits in with those reported by \citet{egc+13}. 

\subsubsection{Irradiation}
\label{sec:irradiation}
We can also investigate the process by which J1023 irradiates its companion. \citet{tct12} suggest that $\gamma$-ray emission from an active radio pulsar is responsible for the heating of the near side of the companion in many systems, and in J1023 in particular. We find that in J1023 the $\gamma$-ray luminosity, if isotropic, is less than that needed to explain the heating of the companion \citep{ta05}. One normally expects a strong pulsar wind consisting of a relativisic pair plasma to carry most of the pulsar's spin-down power. This is therefore a natural candidate to produce irradiation, either directly if it (or heavy ions entrained in it) can reach the companion's surface or indirectly through heating by emission from a shock powered by the pulsar wind. 

In fact, \citet{bah+11} argue that in J1023 there is an X-ray-emitting shock near the first Lagrange point (L1). If it is this shock that provides the X- or $\gamma$-rays that heat the companion, then based on the shock's nearness to the companion's surface we should expect the companion to intercept roughly half of the total X- or $\gamma$-ray luminosity of the shock. If this is the case then both the X- and $\gamma$-ray luminosities are sufficient to explain the companion's heating. While the observed X-ray orbital variability strongly suggests that a substantial fraction of the X-rays come from this region, the lack of detectable $\gamma$-ray variability suggests that only a small fraction of the $\gamma$-rays can come from close to the companion (or we would expect eclipses). Nevertheless the $\gamma$-ray luminosity is sufficient that even if only a small fraction comes from the shock, $\gamma$-rays could be responsible for heating the companion. For the original black widow pulsar, PSR~B1957+20, such an intrabinary shock was predicted by \citet{at93}, and \citet{sgk+03} examined \emph{Chandra X-ray Observatory} data looking for evidence of one. In particular, in line with the predictions, they observe a peak in the X-ray brightness (due to Doppler boosting) immediately following the X-ray eclipse; the same effect was also observed for J1023 by \citet{bah+11}. It therefore seems quite plausible that the companion is being heated by X-rays generated in an intrabinary shock near L1 powered by the particle wind from the pulsar.

In addition to the orbital modulation of the X-ray emission, \citet{akb+10} and \citet{bah+11} note that J1023 shows substantial orbit-to-orbit variability. We do not expect much variability from the MSP in J1023, either in X-ray emission or in particle wind flux. It therefore seems likely that the origin of the orbit-to-orbit X-ray variability comes from the other side of the shock: from the companion. If the amount or magnetization of the material leaving the companion is varying, then an unusually large variation might push material into the pulsar's Roche lobe, form a disc, and tip the system over into another active phase. In light of the fact that the irradiation of the companion --- equal in amplitude to 20\% of the companion's bolometric luminosity --- seems likely to be coming from this shock, it seems that a feed-forward effect might be possible: if an increase in material brightens the shock, the shock will heat the face of the companion more, driving more material off the companion and into the shock. It is equally possible that an increase in material might dim the shock, but nevertheless the shock variability and the corresponding variations in irradiation may well be linked to the system's active phases.

\subsubsection{Ejection}
Energetics can also serve to place an upper bound on how much material can be leaving the system, through evaporation or through Roche lobe overflow followed by expulsion from the system. If the pulsar's entire spin-down luminosity were used to push material out of the system, the companion could be losing $4\times 10^{-7} M_\Sun\;\text{yr}^{-1}$. If instead we assume that exactly the fraction of the pulsar's spin-down luminosity (presumed isotropic) intercepted by the companion is used in ejecting material from the system, we obtain a mass-loss rate of $5\times 10^{-9} M_\Sun\;\text{yr}^{-1}$. This is the typical assumption used when modelling redback and black-widow systems, but it remains somewhat unclear how exactly this wind power would be transformed into ejection of material: it seems that a shock lies between the pulsar and its companion, so that the companion is not directly impacted by the wind; it is not clear how the relatively mild heating experienced by the companion could drive such a powerful outflow.

\subsection{Ionized material in the system}
\label{sec:ionized}

It is clear from the eclipses, shorter signal disappearances, and DM excesses that there is much ionized material in or around the J1023 system. We will discuss three possible models for the nature and location of this material: a magnetically supported wind rather like that of our Sun, a ram-pressure-supported ablative outflow, and freely flowing material overflowing from the companion's Roche lobe. 

The main eclipse is the most consistent feature in our radio observations. It is clear from Figure~\ref{fig:roche} that the material lies far outside the Roche lobe, and in fact the eclipsing material cannot be much closer to the companion than it is to the pulsar. It should also be noted that since the eclipse region is larger at lower frequencies, the density of the eclipsing material must fall off gradually, so that if the eclipse region is bounded by a shock it must encompass the regions eclipsed at all frequencies.

\subsubsection{Magnetically supported material}
If the eclipsing material is trapped in the companion's magnetosphere, the magnetosphere must contain a strong enough magnetic field to resist the pulsar wind. The pulsar wind pressure at the companion's distance (assuming that the pulsar wind power equals the spin-down power, and that the pulsar wind is isotropic) is $7\;\text{dyne}\;\text{cm}^{-2}$; the magnetic field required to produce an equal pressure is $13\;\text{G}$. Can such a field be produced by the companion?

The companion appears to be a G star \citep{ta05}. Since the irradiation produces only $\sim 400\;\text{K}$ difference in temperature between the sides of the companion \citep{ta05}, we expect it to play a fairly minor role in the companion's internal processes. Nevertheless, the companion is presumably not a typical G star: we expect that mass transfer has resulted in it becoming helium-enriched, it has about 1.8 times the radius of a typical $0.24M_\sun$ main-sequence star \citep{fc12}, and the Roche geometry means it is not even spherical. If we assume regardless that the companion has a convective layer and radiative core similar to that of solar-type stars, we should expect a dynamo process to generate a magnetic field. Since the companion is presumably tidally locked, its rotational period is only $0.2\;\text{days}$ (equatorial velocity $110\;\text{km}\;\text{s}^{-1}$), and such rapidly rotating stars tend to have stronger magnetic fields than ordinary solar-type stars. The Sun itself has a magnetic field on the order of $10\;\text{G}\;(R/R_\Sun)^{-2}$ \citep[where $R$ is the distance from the center of the Sun;][]{bam98}. The J1023 system is small enough that the eclipsing regions are within a few solar radii, so it seems plausible that magnetic pressure could support a bubble against the pulsar wind. Relatively tenuous material leaving the surface of the companion would then stream along field lines and might produce the eclipses we observe. Filamentary structures in the companion's magnetosphere might also account for the short-term disappearances of signal at random orbital phases; given their tens-of-seconds duration and the velocity of the pulsar on the plane of the sky, they should be hundreds to thousands of kilometers across. 

Such a magnetically threaded wind should carry angular momentum away from the system, producing magnetic braking and driving continuing accretion. If, on the other hand, the companion is fully convective, we do not expect it to be strongly magnetic \citep{rvj83}, and a magnetically supported wind seems unlikely.

\subsubsection{Ram-pressure-supported material}
We also know that the companions in redbacks and black widows are being irradiated by the pulsars. The companion to PSR~J1023+0038, in particular, fills or nearly fills its Roche lobe, so material is readily removed from its surface, and as we computed in Section~\ref{sec:irradiation}, there is enough power in the part of the pulsar wind it intercepts to remove $5\times 10^{-9}M_\Sun\;\text{yr}^{-1}$ of material. The ram pressure of such a wind might sustain a region in which material from the companion could produce eclipses. \citet{bs13} discuss pulsar and companion winds in redback and other systems, in the interest of understanding the nebulae they produce. They provide a description of the wind balance within the binary system, parameterized by the dimensionless parameter $\eta$, which describes the balance between pulsar and companion winds:
\[
    \eta = \frac{L_p/c}{\dot M v_w},
\]
where $L_p$ is the pulsar wind luminosity, $\dot M$ is the mass loss rate of the companion, and $v_w$ is the velocity with which the companion expels its wind. If we assume a wind from the companion of $5\times 10^{-9}M_\Sun\;\text{yr}^{-1}$, leaving at the companion's escape velocity, we obtain $\eta=0.1$, indicating a system dominated by the companion's wind; the pulsar's wind only fills a region with half-opening angle given by $\psi = 2.1(1-\eta^{2/5})\eta^{1/3}\sim 50\degree$, not very different from the size of the non-eclipsed regions at low frequency, particularly considering that \citet{bah+11} argued that the pulsar wind was concentrated in the plane, leaving the companion's wind more able to expand above the system plane. 

On the other hand, it is not easy to reconcile such a dense wind with the excess DM we measure immediately post-eclipse (and elsewhere): if the wind carries this much mass and moves at this speed, at the distance of the companion's closest approach to our line of sight, the electron density should be of order $10^{16}\;\text{cm}^{-3}$, which implies that the thickness of the layer producing the excess DM must be of the order $300\;\text{km}$. This is wildly inconsistent with the assumption of isotropic outflow used to estimate the electron density; already at the above density the plasma frequency is on the order of $1000\;\text{MHz}$, so increasing the density to produce a thinner layer consistent with the DM measurement would block the $1.4$-GHz radio waves we used to measure the excess DM. If the outflowing material were only partially ionized its effect on DM would be reduced, but it is even more unclear how the pulsar irradiation, by direct wind impact or by reprocessing through an X-ray- or $\gamma$-ray-emitting shock, could drive such a wind without ionizing the material. It therefore seems unclear how the system could have a mass loss rate high enough to hold off the pulsar wind through ram pressure without producing DM and/or eclipses more severe than we currently see.

\subsubsection{Roche-lobe-overflowing material}
There is a third possible source for eclipsing material. In PSR~J1740$-$5340 (NGC~6397A), a globular-cluster redback, the H$\alpha$ line profile suggests  material streaming out from L1, trailing behind the companion \citep{sgf+03}. Since J1023 likely had an accretion disc in 2001, it seems very plausible that some small amount of Roche lobe overflow may be continuing in this system as well. It is, however, difficult to explain the observed eclipse behavior using only such material. While the 1400-MHz eclipse is somewhat asymmetrical, it definitely begins before L1 crosses the line of sight at orbital phase 0.25. The 150 and 350-MHz eclipse regions are even larger and symmetrical. Finally, it is unclear how such escaping material could come to be so far above the plane of the system as to affect our line of sight. 

There remain problems with all these eclipse models. In particular, \citet{bah+11} modelled the X-ray variability as the eclipse of an emitting region by the companion and concluded that most of the X-rays originate from a shocked region near the L1 point or on the pulsar-facing surface of the companion. It is difficult to explain how such a shocked region could be so close to the companion if the eclipsing material, necessarily on the companion side of the shock, reaches as far out as we see in the low-frequency eclipses. \citet{bah+11} argue that the pulsar wind is concentrated in the orbital plane, but it would require very strong beaming to push the shock front down close to the companion's surface within the plane but allow the shock to reach so far around the pulsar above the plane. If the eclipsing material is bound in the star's magnetosphere, this magnetosphere may be anisotropic, and in particular, less extensive around the star's equator. Or, if the eclipsing material is in the form of a wind from the companion, equatorial focusing of the pulsar wind may produce a large $\eta$ in this plane and a much smaller value out of the plane.

\subsection{Orbital variations}
\label{sec:discorbital}

Orbital period variations have been observed in many of the MSP systems that show the effects of ionized material around the system. In the classical black widow systems, this takes the form of slow, quasi-periodic orbital period variations \citep{aft94,lvt+11}. The redbacks and some of the black widows seem to show more complicated orbital period variations (e.g. Hessels et al. 2013, in prep), and \citet{bkr+13} suggested that the systems whose companions have higher Roche-lobe filling factors also show stronger orbital period variations. The accreting millisecond X-ray pulsar SAX~J1808.4$-$3658, which is thought to contain a (not yet detected) radio pulsar in quiescence \citep{bdd+03,ibb+09}, also shows evidence for a non-zero second derivative of orbital period, possibly indicating that it too undergoes orbital period variations \citep{pbg+12}. J1023 fits into this picture, since it appears to have a high Roche-lobe filling factor and also substantial orbital period variations.

\subsubsection{Gravitational Quadrupole Coupling}
One specific mechanism, gravitational quadrupole coupling (GQC), by which motions within the companion might be coupled to the orbital motion, was set out by \citet{appl92} in order to understand close optical binaries such as Algol. The model was applied to the PSR~B1957+20 system (the original black widow) by \citet{as94}. In this model, as a result of magnetic braking, the companion would have a surface layer rotating at a different speed from the underlying layers. The magnetic field within the companion would produce a torque coupling the layers, and variations in this magnetic field would then vary the rotation of the outer layer; if the outer layer spins more rapidly, the star becomes more oblate, while if it slows, the star becomes less oblate. This change in the mass distribution within the companion, specifically in the quadrupole moment of the mass distribution, would then change the orbital period. Normally, such changes in the differential rotation require substantial energy transfer, and \citet{appl92} argues that this requires a corresponding fractional change in luminosity of the star of about $10\%$. Such variations are observed in a number of non-degenerate binary systems, and the luminosity variations are correlated in time with the orbital period variations as predicted by the model \citep{appl92}.

In J1023, we see orbital phase variations on the order of $1\,\text{s}$. The apparent random wander in orbital period seems to produce this variation on the timescale of a few months, though we see no evidence for the quasi-periodicity described in \citet{appl92} and seen in many systems with orbital period variations \citep[for example][]{appl92,aft94,lvt+11}. The lack of visible periodicity could be due to a timescale shorter than we can resolve, so that we would see only the net result of more than one quasi-periodic cycle. However, applying the calculations below to such a short modulation period predicts a tremendous energy flow within the companion, implying luminosity variations over a hundred times the bolometric luminosity of the companion.  In \citet{appl92} the measured parameters of the source RS~CVn imply luminosity variations almost ten times the luminosity of the star, so deviations from the model are known to occur. Nevertheless it seems implausible in this model that the timescale could be shorter than our observation cadence so as to hide the quasi-periodicity.

For the calculations that follow we will assume a time scale for the modulation of $200$ days, plausible in light of Figure~\ref{fig:orbvariations} but chosen to yield luminosity variations consistent with observational constraints. We will also assume, a necessity in this model, that the variations are quasi-periodic with period $P_{\text{mod}}$. The amplitude of the phase variations $\Delta t\sim1\,\text{s}$ implies fractional orbital period variations $\Delta P/P$ on the order of $2\pi\Delta t/P_{\text{mod}}=4\times 10^{-7}$. Following \citet{appl92} and \citet{as94}, we assume that the participating shell contains about a tenth of the companion's mass. They link the orbital period variations to changes in the quadrupole moment $\Delta Q$ by
$$
\frac{\Delta P}{P} = -9\left(\frac{R_c}{a}\right)^2\frac{\Delta Q}{M_c R_c^2},
$$
where $M_c$, $R_c$ and $a$ are the mass and radius of the companion and the semimajor axis of the orbit, respectively. Changes in the quadrupole moment are dominated by changes in the rotation rate of the outer shell, so they write
$$
\Delta Q = \frac{2}{9} \Omega\Delta\Omega\frac{M_s R_c^5}{GM_c},
$$
where $M_s$ is the mass of the shell and $\Omega$ is its angular velocity.
Using this we infer a fractional change in the shell's angular velocity on the order of $2\times 10^{-4}$. Again following the above works, we assume that the companion is slowed below corotation by the same fractional amount, $2\times 10^{-4}$. In this model, the angular momentum transfer between layers is produced by a torque coming from the subsurface magnetic field, acting quasi-periodically over the modulation period $P_{\text{mod}}$. They estimate the strength of this subsurface magnetic field $B_s$ as
$$
B_s^2 \sim 10\frac{GM_c^2}{R_c^4}\left(\frac{a}{R_c}\right)^2\frac{\Delta P}{P_{\text{mod}}}.
$$ 
In J1023 this calculation finds $B_s\sim 4\times 10^4\,\text{G}$, greater by an order of magnitude than those in the systems considered in \citet{appl92}. While this subsurface field is difficult to constrain observationally, most of the systems considered by \citet{appl92} and \citet{as94} have values of this and other parameters that are surprisingly similar to each other, so such a large value for J1023 is unusual, and difficult to explain if the companion has become non-magnetic. On the other hand, the observations summarized in \citet{ta05} constrain luminosity variations to less than about $0.04$ magnitudes, which is about $4\%$. The GQC model predicts that the transfer of kinetic energy between layers should be accompanied by luminosity variations on the same scale, in this case about $1.5\times 10^{31}\,\text{erg}\,\text{s}^{-1}$; by our selection of $P_{\text{mod}}$ this is about $2\%$ of the companion's luminosity, below the observational limits on such variations. Nevertheless, this low upper limit is a little puzzling in this model, since it implies that the companion is less efficient than usual at converting differential rotation into luminosity, in spite of a higher-than-normal subsurface magnetic field. Most systems discussed in \citet{appl92} undergo luminosity variations of more than about $10\%$, which would have been detected by \citet{ta05} unless they were on a much longer timescale than the orbital period variations. 

\subsubsection{Relation to active phases}
If the orbital period variations are connected to mass motions within the companion, whether through the mechanism of \citet{appl92} or some other process, they may be connected to J1023's active phases. In particular, if the star very nearly fills its Roche lobe, shape distortions, whether global or local, might easily cause substantial amounts of material to spill out of the Roche lobe through L1. Returning to the GQC model, the fractional variations in the equatorial radius we expect are on the order of $10^{-4}$ (proportional to the fractional change in the shell's angular velocity). Substantial enough amounts of material spilling through L1 can overcome the pulsar wind pressure and penetrate the light cylinder; once this occurs, the process by which the wind is generated becomes ineffective, and an accretion disc can continue to exist even if the rate of mass transfer drops somewhat. This bistable mechanism, proposed by \citet{bdd+03}, could be made to switch states by random or ordered changes within the companion.

The GQC mechanism of \citet{appl92} requires magnetic braking to slow the companion below corotation, and it also requires relatively strong magnetic fields within the companion. More generally, it seems difficult to produce orbital period variations without the companion being slowed below corotation. This slowing would seem to imply loss of angular momentum from the binary. If this is occurring, then the Roche lobe must be shrinking. If a system has a main-sequence companion that is close to filling its Roche lobe, then it seems that orbital period variations should imply that Roche lobe overflow is continuing unless there is evaporation sufficient to outpace the Roche lobe shrinkage.

\subsection{Active phases}

In summary, the presence of ionized material in the system, the Roche-lobe-filling companion, the irradiation of the companion by a shock near L1, and the orbital period variations all seem to be linked to the active phases. Many of these phenomena have been seen in other systems; the class of redbacks is largely defined by these phenomena, and redbacks are being discovered in increasing numbers \citep{hrm+11,rrc+13}. And yet if these objects regularly undergo accretion, why have we not detected many more LMXBs? A normal LMXB at the usual distance for these redbacks would certainly have been detected by all-sky monitors. The answer to this question may be hinted at by J1023's active episode: J1023 never became very bright during its active phase, in spite of evidence for the presence of a disc. \citet{asr+09} suggested that J1023 might have undergone ``propeller-mode accretion,'' in which material is ejected rather than accreting onto the neutron-star surface, but other possibilities exist. There are also emerging categories of faint and very faint LMXBs, which never seem to become very bright during their outbursts. So while we think it likely that sources with Roche-lobe-filling companions, ionized material in the system, and orbital period variations --- in short, redbacks --- will undergo active phases, and while the active phase of PSR~J1824$-$2452I was X-ray bright, these active phases may in many cases be quite X-ray faint. 

\section{Conclusions}

J1023 exhibits a variety of observational phenomena: eclipses, excess dispersion measure, short-term disappearances of signal, and orbital period variations. The origin of these phenomena is not clear, but a magnetically active companion may be able to explain several: the eclipses may be the result of ionized material trapped in a bubble supported against the pulsar wind by the companion's magnetic field. The excess dispersion measure and disappearances of signal may be due to blobs or tendrils of magnetically trapped material extending outside the primary bubble in the manner of solar prominences. Finally, the orbital period variations may be the result of the mechanism of \citet{appl92}, in which this same magnetic field slows the companion slightly below corotation, and in which magnetic fields internal to the companion transfer angular momentum to and from the surface layers, modifying the companion's quadrupole moment and thus the system's orbital period. A magnetically active companion is plausible in light of the fact that the companion resembles a main-sequence star and, with a rotational period approximately equal to the system's $0.198$-day orbit, the companion is also a rapid rotator. That said, a magnetically active companion would presumably still be undergoing magnetic braking, which should drive Roche lobe overflow and hence active accretion: it is difficult to explain why such a system would not appear as a normal accreting X-ray binary. In fact, models of binary evolution predict that at companion masses of ${\sim}0.3 M_\Sun$ the companion should become fully convective and lose its magnetic field. In J1023 we find it difficult to reconcile a non-magnetic companion with the eclipses, observed modest density of the outflowing material, and orbital period variations. Magnetic braking is a very poorly understood process, so perhaps detailed study of J1023's current state may clarify the role of magnetic braking in binary evolution, and may even allow direct study of a system undergoing magnetic braking. 

There are several promising avenues for further study of J1023. Careful optical observations to check for luminosity variations in the companion would help clarify whether the GQC mechanism can plausibly be acting in the system, and if not, might help clarify other possible mechanisms. Multicolor light curve modeling and possibly orbital-phase-resolved optical spectroscopy might clarify how close the companion is to filling its Roche lobe, and asymmetries in the light curve might give some indication of the mechanism by which the companion is irradiated. Further X-ray observations, particularly if high time resolution is available, might make it possible to distinguish X-rays from the pulsar itself from X-rays originating in a shock region, possibly even revealing the shock geometry. 

Now that J1023 appears to have entered another active phase, we anticipate active research into its unusual accretion state. The description of J1023 in its radio pulsar state that we provide should provide a detailed picture of the system geometry and a solid basis for comparison with the system's behaviour in the active state. When the active phase ends and the radio pulsar again becomes active, timing the pulsar should provide extremely valuable constraints on the accretion models: propeller-mode accretion, for example, is expected to slow the pulsar down, while a state in which the light cylinder remains clear should not. Close monitoring of this very unusual system should clarify our picture of the last stages of accretion and the birth of a millisecond pulsar.

\acknowledgements
The authors would like to thank Chris Thompson, Joanna Rankin, and Alessandro Patruno for productive discussions. The authors would also like to thank the Green Bank North Celestial Cap pulsar survey team for allowing us to observe J1023 at 350~MHz, doubling as validation scans for the survey. VMK aknowledges support from an NSERC Discovery Grant and Accelerator Supplement, from the Canadian Institute for Advanced Research, the Canada Research Chairs Program, FQRNT, and the Lorne Trottier Chair in Astrophysics and Cosmology. JWTH acknowledges funding through the Netherlands Foundation for Scientific research (NWO) and a Starting Grant from the European Research Council (ERC). The \textit{Fermi} LAT Collaboration acknowledges generous ongoing support from a number of agencies and institutes that have supported both the development and the operation of the LAT as well as scientific data analysis. These include the National Aeronautics and Space Administration and the Department of Energy in the United States, the Commissariat \`a l'Energie Atomique and the Centre National de la Recherche Scientifique / Institut National de Physique Nucl\'eaire et de Physique des Particules in France, the Agenzia Spaziale Italiana and the Istituto Nazionale di Fisica Nucleare in Italy, the Ministry of Education, Culture, Sports, Science and Technology (MEXT), High Energy Accelerator Research Organization (KEK) and Japan Aerospace Exploration Agency (JAXA) in Japan, and the K.~A.~Wallenberg Foundation, the Swedish Research Council and the Swedish National Space Board in Sweden. Additional support for science analysis during the operations phase is gratefully acknowledged from the Istituto Nazionale di Astrofisica in Italy and the Centre National d'\'Etudes Spatiales in France. The Westerbork Synthesis Radio Telescope is operated by ASTRON, the Netherlands Institute for Radio Astronomy, with support from NWO. Access to the Lovell Telescope is supported through an STFC consolidated grant. The Arecibo Observatory is operated by SRI International under a cooperative agreement with the National Science Foundation (AST-1100968), and in alliance with Ana G. Méndez-Universidad Metropolitana, and the Universities Space Research Association. The National Radio Astronomy Observatory is a facility of the National Science Foundation operated under cooperative agreement by Associated Universities, Inc.

\emph{Facilities:} \emph{\facility{Arecibo}}, \emph{\facility{Fermi}}, \emph{\facility{GBT}}, \emph{\facility{WSRT}}, \emph{\facility{Lovell}} 

\bibliographystyle{apj}
\bibliography{journals,refs}

\begin{thebibliography}{60}
\expandafter\ifx\csname natexlab\endcsname\relax\def\natexlab#1{#1}\fi

\bibitem[{{Abdo} {et~al.}(2013){Abdo}, {Ajello}, {Allafort}, {Baldini},
  {Ballet}, {Barbiellini}, {Baring}, {Bastieri}, {Belfiore}, {Bellazzini}, \&
  et~al.}]{aaa+13}
{Abdo}, A.~A., {Ajello}, M., {Allafort}, A., {Baldini}, L., {Ballet}, J.,
  {Barbiellini}, G., {Baring}, M.~G., {Bastieri}, D., {Belfiore}, A.,
  {Bellazzini}, R., \& et~al. 2013, \apjs, 208, 17

\bibitem[{{Ackermann} {et~al.}(2012){Ackermann}, {Ajello}, {Albert},
  {Allafort}, {Atwood}, {Axelsson}, {Baldini}, {Ballet}, {Barbiellini},
  {Bastieri}, {Bechtol}, {Bellazzini}, {Bissaldi}, {Blandford}, {Bloom},
  {Bogart}, {Bonamente}, {Borgland}, {Bottacini}, {Bouvier}, {Brandt},
  {Bregeon}, {Brigida}, {Bruel}, {Buehler}, {Burnett}, {Buson}, {Caliandro},
  {Cameron}, {Caraveo}, {Casandjian}, {Cavazzuti}, {Cecchi}, {{\c C}elik},
  {Charles}, {Chaves}, {Chekhtman}, {Cheung}, {Chiang}, {Ciprini}, {Claus},
  {Cohen-Tanugi}, {Conrad}, {Corbet}, {Cutini}, {D'Ammando}, {Davis}, {de
  Angelis}, {DeKlotz}, {de Palma}, {Dermer}, {Digel}, {Silva}, {Drell},
  {Drlica-Wagner}, {Dubois}, {Favuzzi}, {Fegan}, {Ferrara}, {Focke}, {Fortin},
  {Fukazawa}, {Funk}, {Fusco}, {Gargano}, {Gasparrini}, {Gehrels}, {Giebels},
  {Giglietto}, {Giordano}, {Giroletti}, {Glanzman}, {Godfrey}, {Grenier},
  {Grove}, {Guiriec}, {Hadasch}, {Hayashida}, {Hays}, {Horan}, {Hou}, {Hughes},
  {Jackson}, {Jogler}, {J{\'o}hannesson}, {Johnson}, {Johnson}, {Johnson},
  {Kamae}, {Katagiri}, {Kataoka}, {Kerr}, {Kn{\"o}dlseder}, {Kuss}, {Lande},
  {Larsson}, {Latronico}, {Lavalley}, {Lemoine-Goumard}, {Longo}, {Loparco},
  {Lott}, {Lovellette}, {Lubrano}, {Mazziotta}, {McConville}, {McEnery},
  {Mehault}, {Michelson}, {Mitthumsiri}, {Mizuno}, {Moiseev}, {Monte},
  {Monzani}, {Morselli}, {Moskalenko}, {Murgia}, {Naumann-Godo}, {Nemmen},
  {Nishino}, {Norris}, {Nuss}, {Ohno}, {Ohsugi}, {Okumura}, {Omodei},
  {Orienti}, {Orlando}, {Ormes}, {Paneque}, {Panetta}, {Perkins},
  {Pesce-Rollins}, {Pierbattista}, {Piron}, {Pivato}, {Porter}, {Racusin},
  {Rain{\`o}}, {Rando}, {Razzano}, {Razzaque}, {Reimer}, {Reimer}, {Reposeur},
  {Reyes}, {Ritz}, {Rochester}, {Romoli}, {Roth}, {Sadrozinski}, {Sanchez},
  {Saz Parkinson}, {Sbarra}, {Scargle}, {Sgr{\`o}}, {Siegal-Gaskins},
  {Siskind}, {Spandre}, {Spinelli}, {Stephens}, {Suson}, {Tajima}, {Takahashi},
  {Tanaka}, {Thayer}, {Thayer}, {Thompson}, {Tibaldo}, {Tinivella}, {Tosti},
  {Troja}, {Usher}, {Vandenbroucke}, {Van Klaveren}, {Vasileiou}, {Vianello},
  {Vitale}, {Waite}, {Wallace}, {Winer}, {Wood}, {Wood}, {Wood}, {Yang}, \&
  {Zimmer}}]{aaa+12}
{Ackermann}, M., {Ajello}, M., {Albert}, A., {Allafort}, A., {Atwood}, W.~B.,
  {Axelsson}, M., {Baldini}, L., {Ballet}, J., {Barbiellini}, G., {Bastieri},
  D., {Bechtol}, K., {Bellazzini}, R., {Bissaldi}, E., {Blandford}, R.~D.,
  {Bloom}, E.~D., {Bogart}, J.~R., {Bonamente}, E., {Borgland}, A.~W.,
  {Bottacini}, E., {Bouvier}, A., {Brandt}, T.~J., {Bregeon}, J., {Brigida},
  M., {Bruel}, P., {Buehler}, R., {Burnett}, T.~H., {Buson}, S., {Caliandro},
  G.~A., {Cameron}, R.~A., {Caraveo}, P.~A., {Casandjian}, J.~M., {Cavazzuti},
  E., {Cecchi}, C., {{\c C}elik}, {\"O}., {Charles}, E., {Chaves}, R.~C.~G.,
  {Chekhtman}, A., {Cheung}, C.~C., {Chiang}, J., {Ciprini}, S., {Claus}, R.,
  {Cohen-Tanugi}, J., {Conrad}, J., {Corbet}, R., {Cutini}, S., {D'Ammando},
  F., {Davis}, D.~S., {de Angelis}, A., {DeKlotz}, M., {de Palma}, F.,
  {Dermer}, C.~D., {Digel}, S.~W., {Silva}, E.~d.~C.~e., {Drell}, P.~S.,
  {Drlica-Wagner}, A., {Dubois}, R., {Favuzzi}, C., {Fegan}, S.~J., {Ferrara},
  E.~C., {Focke}, W.~B., {Fortin}, P., {Fukazawa}, Y., {Funk}, S., {Fusco}, P.,
  {Gargano}, F., {Gasparrini}, D., {Gehrels}, N., {Giebels}, B., {Giglietto},
  N., {Giordano}, F., {Giroletti}, M., {Glanzman}, T., {Godfrey}, G.,
  {Grenier}, I.~A., {Grove}, J.~E., {Guiriec}, S., {Hadasch}, D., {Hayashida},
  M., {Hays}, E., {Horan}, D., {Hou}, X., {Hughes}, R.~E., {Jackson}, M.~S.,
  {Jogler}, T., {J{\'o}hannesson}, G., {Johnson}, R.~P., {Johnson}, T.~J.,
  {Johnson}, W.~N., {Kamae}, T., {Katagiri}, H., {Kataoka}, J., {Kerr}, M.,
  {Kn{\"o}dlseder}, J., {Kuss}, M., {Lande}, J., {Larsson}, S., {Latronico},
  L., {Lavalley}, C., {Lemoine-Goumard}, M., {Longo}, F., {Loparco}, F.,
  {Lott}, B., {Lovellette}, M.~N., {Lubrano}, P., {Mazziotta}, M.~N.,
  {McConville}, W., {McEnery}, J.~E., {Mehault}, J., {Michelson}, P.~F.,
  {Mitthumsiri}, W., {Mizuno}, T., {Moiseev}, A.~A., {Monte}, C., {Monzani},
  M.~E., {Morselli}, A., {Moskalenko}, I.~V., {Murgia}, S., {Naumann-Godo}, M.,
  {Nemmen}, R., {Nishino}, S., {Norris}, J.~P., {Nuss}, E., {Ohno}, M.,
  {Ohsugi}, T., {Okumura}, A., {Omodei}, N., {Orienti}, M., {Orlando}, E.,
  {Ormes}, J.~F., {Paneque}, D., {Panetta}, J.~H., {Perkins}, J.~S.,
  {Pesce-Rollins}, M., {Pierbattista}, M., {Piron}, F., {Pivato}, G., {Porter},
  T.~A., {Racusin}, J.~L., {Rain{\`o}}, S., {Rando}, R., {Razzano}, M.,
  {Razzaque}, S., {Reimer}, A., {Reimer}, O., {Reposeur}, T., {Reyes}, L.~C.,
  {Ritz}, S., {Rochester}, L.~S., {Romoli}, C., {Roth}, M., {Sadrozinski},
  H.~F.-W., {Sanchez}, D.~A., {Saz Parkinson}, P.~M., {Sbarra}, C., {Scargle},
  J.~D., {Sgr{\`o}}, C., {Siegal-Gaskins}, J., {Siskind}, E.~J., {Spandre}, G.,
  {Spinelli}, P., {Stephens}, T.~E., {Suson}, D.~J., {Tajima}, H., {Takahashi},
  H., {Tanaka}, T., {Thayer}, J.~G., {Thayer}, J.~B., {Thompson}, D.~J.,
  {Tibaldo}, L., {Tinivella}, M., {Tosti}, G., {Troja}, E., {Usher}, T.~L.,
  {Vandenbroucke}, J., {Van Klaveren}, B., {Vasileiou}, V., {Vianello}, G.,
  {Vitale}, V., {Waite}, A.~P., {Wallace}, E., {Winer}, B.~L., {Wood}, D.~L.,
  {Wood}, K.~S., {Wood}, M., {Yang}, Z., \& {Zimmer}, S. 2012, \apjs, 203, 4

\bibitem[{{Applegate}(1992)}]{appl92}
{Applegate}, J.~H. 1992, \apj, 385, 621

\bibitem[{{Applegate} \& {Shaham}(1994)}]{as94}
{Applegate}, J.~H. \& {Shaham}, J. 1994, \apj, 436, 312

\bibitem[{{Archibald} {et~al.}(2010){Archibald}, {Kaspi}, {Bogdanov},
  {Hessels}, {Stairs}, {Ransom}, \& {McLaughlin}}]{akb+10}
{Archibald}, A.~M., {Kaspi}, V.~M., {Bogdanov}, S., {Hessels}, J.~W.~T.,
  {Stairs}, I.~H., {Ransom}, S.~M., \& {McLaughlin}, M.~A. 2010, \apj, 722, 88

\bibitem[{{Archibald} {et~al.}(2009){Archibald}, {Stairs}, {Ransom}, {Kaspi},
  {Kondratiev}, {Lorimer}, {McLaughlin}, {Boyles}, {Hessels}, {Lynch}, {van
  Leeuwen}, {Roberts}, {Jenet}, {Champion}, {Rosen}, {Barlow}, {Dunlap}, \&
  {Remillard}}]{asr+09}
{Archibald}, A.~M., {Stairs}, I.~H., {Ransom}, S.~M., {Kaspi}, V.~M.,
  {Kondratiev}, V.~I., {Lorimer}, D.~R., {McLaughlin}, M.~A., {Boyles}, J.,
  {Hessels}, J.~W.~T., {Lynch}, R., {van Leeuwen}, J., {Roberts}, M.~S.~E.,
  {Jenet}, F., {Champion}, D.~J., {Rosen}, R., {Barlow}, B.~N., {Dunlap},
  B.~H., \& {Remillard}, R.~A. 2009, Science, 324, 1411

\bibitem[{{Arons} \& {Tavani}(1993)}]{at93}
{Arons}, J. \& {Tavani}, M. 1993, \apj, 403, 249

\bibitem[{{Arzoumanian} {et~al.}(1994){Arzoumanian}, {Fruchter}, \&
  {Taylor}}]{aft94}
{Arzoumanian}, Z., {Fruchter}, A.~S., \& {Taylor}, J.~H. 1994, \apjl, 426, L85

\bibitem[{{Atwood} {et~al.}(2009){Atwood}, {Abdo}, {Ackermann}, {Althouse},
  {Anderson}, {Axelsson}, {Baldini}, {Ballet}, {Band}, {Barbiellini}, \&
  et~al.}]{aaa+09}
{Atwood}, W.~B., {Abdo}, A.~A., {Ackermann}, M., {Althouse}, W., {Anderson},
  B., {Axelsson}, M., {Baldini}, L., {Ballet}, J., {Band}, D.~L.,
  {Barbiellini}, G., \& et~al. 2009, \apj, 697, 1071

\bibitem[{{Banaszkiewicz} {et~al.}(1998){Banaszkiewicz}, {Axford}, \&
  {McKenzie}}]{bam98}
{Banaszkiewicz}, M., {Axford}, W.~I., \& {McKenzie}, J.~F. 1998, \aap, 337, 940

\bibitem[{{Bednarek} \& {Sitarek}(2013)}]{bs13}
{Bednarek}, W. \& {Sitarek}, J. 2013, \mnras, 430, 2951

\bibitem[{{Bogdanov} {et~al.}(2011){Bogdanov}, {Archibald}, {Hessels}, {Kaspi},
  {Lorimer}, {McLaughlin}, {Ransom}, \& {Stairs}}]{bah+11}
{Bogdanov}, S., {Archibald}, A.~M., {Hessels}, J.~W.~T., {Kaspi}, V.~M.,
  {Lorimer}, D., {McLaughlin}, M.~A., {Ransom}, S.~M., \& {Stairs}, I.~H. 2011,
  \apj, 742, 97

\bibitem[{{Breton} {et~al.}(2013){Breton}, {van Kerkwijk}, {Roberts},
  {Hessels}, {Camilo}, {McLaughlin}, {Ransom}, {Ray}, \& {Stairs}}]{bkr+13}
{Breton}, R.~P., {van Kerkwijk}, M.~H., {Roberts}, M.~S.~E., {Hessels},
  J.~W.~T., {Camilo}, F., {McLaughlin}, M.~A., {Ransom}, S.~M., {Ray}, P.~S.,
  \& {Stairs}, I.~H. 2013, ArXiv e-prints

\bibitem[{{B{\"u}ning} \& {Ritter}(2004)}]{br04}
{B{\"u}ning}, A. \& {Ritter}, H. 2004, \aap, 423, 281

\bibitem[{{Burderi} {et~al.}(2003){Burderi}, {Di Salvo}, {D'Antona}, {Robba},
  \& {Testa}}]{bdd+03}
{Burderi}, L., {Di Salvo}, T., {D'Antona}, F., {Robba}, N.~R., \& {Testa}, V.
  2003, 404, L43

\bibitem[{{Burderi} {et~al.}(2001){Burderi}, {Possenti}, {D'Antona}, {Di
  Salvo}, {Burgay}, {Stella}, {Menna}, {Iaria}, {Campana}, \&
  {d'Amico}}]{bpd+01}
{Burderi}, L., {Possenti}, A., {D'Antona}, F., {Di Salvo}, T., {Burgay}, M.,
  {Stella}, L., {Menna}, M.~T., {Iaria}, R., {Campana}, S., \& {d'Amico}, N.
  2001, \apjl, 560, L71

\bibitem[{{Chen} {et~al.}(2013){Chen}, {Chen}, {Tauris}, \& {Han}}]{ccth13}
{Chen}, H.-L., {Chen}, X., {Tauris}, T.~M., \& {Han}, Z. 2013, \apj, 775, 27

\bibitem[{{Cordes}(1986)}]{cord86}
{Cordes}, J.~M. 1986, \apj, 311, 183

\bibitem[{{Cordes} \& {Rickett}(1998)}]{cr98}
{Cordes}, J.~M. \& {Rickett}, B.~J. 1998, \apj, 507, 846

\bibitem[{{Deller} {et~al.}(2012){Deller}, {Archibald}, {Brisken},
  {Chatterjee}, {Janssen}, {Kaspi}, {Lorimer}, {Lyne}, {McLaughlin}, {Ransom},
  {Stairs}, \& {Stappers}}]{das+12}
{Deller}, A.~T., {Archibald}, A.~M., {Brisken}, W.~F., {Chatterjee}, S.,
  {Janssen}, G.~H., {Kaspi}, V.~M., {Lorimer}, D., {Lyne}, A.~G., {McLaughlin},
  M.~A., {Ransom}, S., {Stairs}, I.~H., \& {Stappers}, B. 2012, \apjl, 756, L25

\bibitem[{{Demorest}(2007)}]{demo07}
{Demorest}, P.~B. 2007, PhD thesis, University of California, Berkeley

\bibitem[{{Doroshenko} {et~al.}(2001){Doroshenko}, {L{\"o}hmer}, {Kramer},
  {Jessner}, {Wielebinski}, {Lyne}, \& {Lange}}]{dlk+01}
{Doroshenko}, O., {L{\"o}hmer}, O., {Kramer}, M., {Jessner}, A., {Wielebinski},
  R., {Lyne}, A.~G., \& {Lange}, C. 2001, 379, 579

\bibitem[{{Dowd} {et~al.}(2000){Dowd}, {Sisk}, \& {Hagen}}]{dsh00}
{Dowd}, A., {Sisk}, W., \& {Hagen}, J. 2000, in Astronomical Society of the
  Pacific Conference Series, Vol. 202, IAU Colloq. 177: Pulsar Astronomy - 2000
  and Beyond, ed. M.~{Kramer}, N.~{Wex}, \& R.~{Wielebinski}, 275

\bibitem[{{Dubus} {et~al.}(1999){Dubus}, {Lasota}, {Hameury}, \&
  {Charles}}]{dlhc99}
{Dubus}, G., {Lasota}, J.-P., {Hameury}, J.-M., \& {Charles}, P. 1999, \mnras,
  303, 139

\bibitem[{{Espinoza} {et~al.}(2013){Espinoza}, {Guillemot}, {{\c C}elik},
  {Weltevrede}, {Stappers}, {Smith}, {Kerr}, {Zavlin}, {Cognard}, {Eatough},
  {Freire}, {Janssen}, {Camilo}, {Desvignes}, {Hewitt}, {Hou}, {Johnston},
  {Keith}, {Kramer}, {Lyne}, {Manchester}, {Ransom}, {Ray}, {Shannon},
  {Theureau}, \& {Webb}}]{egc+13}
{Espinoza}, C.~M., {Guillemot}, L., {{\c C}elik}, {\"O}., {Weltevrede}, P.,
  {Stappers}, B.~W., {Smith}, D.~A., {Kerr}, M., {Zavlin}, V.~E., {Cognard},
  I., {Eatough}, R.~P., {Freire}, P.~C.~C., {Janssen}, G.~H., {Camilo}, F.,
  {Desvignes}, G., {Hewitt}, J.~W., {Hou}, X., {Johnston}, S., {Keith}, M.,
  {Kramer}, M., {Lyne}, A., {Manchester}, R.~N., {Ransom}, S.~M., {Ray}, P.~S.,
  {Shannon}, R., {Theureau}, G., \& {Webb}, N. 2013, \mnras, 430, 571

\bibitem[{{Feiden} \& {Chaboyer}(2012)}]{fc12}
{Feiden}, G.~A. \& {Chaboyer}, B. 2012, \apj, 757, 42

\bibitem[{{Ferdman}(2008)}]{ferd08}
{Ferdman}, R.~D. 2008, PhD thesis, University of British Columbia

\bibitem[{{Fruchter} {et~al.}(1988){Fruchter}, {Stinebring}, \&
  {Taylor}}]{fst88}
{Fruchter}, A.~S., {Stinebring}, D.~R., \& {Taylor}, J.~H. 1988, \nat, 333, 237

\bibitem[{{Guillemot} {et~al.}(2012){Guillemot}, {Johnson}, {Venter}, {Kerr},
  {Pancrazi}, {Livingstone}, {Janssen}, {Jaroenjittichai}, {Kramer}, {Cognard},
  {Stappers}, {Harding}, {Camilo}, {Espinoza}, {Freire}, {Gargano}, {Grove},
  {Johnston}, {Michelson}, {Noutsos}, {Parent}, {Ransom}, {Ray}, {Shannon},
  {Smith}, {Theureau}, {Thorsett}, \& {Webb}}]{gjv+12}
{Guillemot}, L., {Johnson}, T.~J., {Venter}, C., {Kerr}, M., {Pancrazi}, B.,
  {Livingstone}, M., {Janssen}, G.~H., {Jaroenjittichai}, P., {Kramer}, M.,
  {Cognard}, I., {Stappers}, B.~W., {Harding}, A.~K., {Camilo}, F., {Espinoza},
  C.~M., {Freire}, P.~C.~C., {Gargano}, F., {Grove}, J.~E., {Johnston}, S.,
  {Michelson}, P.~F., {Noutsos}, A., {Parent}, D., {Ransom}, S.~M., {Ray},
  P.~S., {Shannon}, R., {Smith}, D.~A., {Theureau}, G., {Thorsett}, S.~E., \&
  {Webb}, N. 2012, \apj, 744, 33

\bibitem[{{Hessels} {et~al.}(2011){Hessels}, {Roberts}, {McLaughlin}, {Ray},
  {Bangale}, {Ransom}, {Kerr}, {Camilo}, \& {Decesar}}]{hrm+11}
{Hessels}, J.~W.~T., {Roberts}, M.~S.~E., {McLaughlin}, M.~A., {Ray}, P.~S.,
  {Bangale}, P., {Ransom}, S.~M., {Kerr}, M., {Camilo}, F., \& {Decesar}, M.~E.
  2011, in American Institute of Physics Conference Series, Vol. 1357, American
  Institute of Physics Conference Series, ed. M.~{Burgay}, N.~{D'Amico},
  P.~{Esposito}, A.~{Pellizzoni}, \& A.~{Possenti}, 40--43

\bibitem[{{Hobbs} {et~al.}(2006){Hobbs}, {Edwards}, \& {Manchester}}]{hem06}
{Hobbs}, G.~B., {Edwards}, R.~T., \& {Manchester}, R.~N. 2006, \mnras, 369, 655

\bibitem[{{Hotan} {et~al.}(2004){Hotan}, {van Straten}, \&
  {Manchester}}]{hsm04}
{Hotan}, A.~W., {van Straten}, W., \& {Manchester}, R.~N. 2004, PASA, 21, 302

\bibitem[{{Iacolina} {et~al.}(2009){Iacolina}, {Burgay}, {Burderi}, {Possenti},
  \& {di Salvo}}]{ibb+09}
{Iacolina}, M.~N., {Burgay}, M., {Burderi}, L., {Possenti}, A., \& {di Salvo},
  T. 2009, \aap, 497, 445

\bibitem[{{Karuppusamy} {et~al.}(2008){Karuppusamy}, {Stappers}, \& {van
  Straten}}]{kss08}
{Karuppusamy}, R., {Stappers}, B., \& {van Straten}, W. 2008, \pasp, 120, 191

\bibitem[{{Kerr}(2011)}]{kerr11}
{Kerr}, M. 2011, \apj, 732, 38

\bibitem[{{Lattimer} \& {Prakash}(2004)}]{lp04}
{Lattimer}, J.~M. \& {Prakash}, M. 2004, Science, 304, 536

\bibitem[{{Lazaridis} {et~al.}(2011){Lazaridis}, {Verbiest}, {Tauris},
  {Stappers}, {Kramer}, {Wex}, {Jessner}, {Cognard}, {Desvignes}, {Janssen},
  {Purver}, {Theureau}, {Bassa}, \& {Smits}}]{lvt+11}
{Lazaridis}, K., {Verbiest}, J.~P.~W., {Tauris}, T.~M., {Stappers}, B.~W.,
  {Kramer}, M., {Wex}, N., {Jessner}, A., {Cognard}, I., {Desvignes}, G.,
  {Janssen}, G.~H., {Purver}, M.~B., {Theureau}, G., {Bassa}, C.~G., \&
  {Smits}, R. 2011, \mnras, 414, 3134

\bibitem[{{Lorimer} \& {Kramer}(2004)}]{lk04}
{Lorimer}, D.~R. \& {Kramer}, M. 2004, {Handbook of Pulsar Astronomy}, ed.
  R.~{Ellis}, J.~{Huchra}, S.~{Kahn}, G.~{Rieke}, \& P.~B. {Stetson}

\bibitem[{{Nice} {et~al.}(1990){Nice}, {Thorsett}, {Taylor}, \&
  {Fruchter}}]{nttf90}
{Nice}, D.~J., {Thorsett}, S.~E., {Taylor}, J.~H., \& {Fruchter}, A.~S. 1990,
  \apjl, 361, L61

\bibitem[{{Nolan} {et~al.}(2012){Nolan}, {Abdo}, {Ackermann}, {Ajello},
  {Allafort}, {Antolini}, {Atwood}, {Axelsson}, {Baldini}, {Ballet}, \&
  et~al.}]{naa+12}
{Nolan}, P.~L., {Abdo}, A.~A., {Ackermann}, M., {Ajello}, M., {Allafort}, A.,
  {Antolini}, E., {Atwood}, W.~B., {Axelsson}, M., {Baldini}, L., {Ballet}, J.,
  \& et~al. 2012, \apjs, 199, 31

\bibitem[{{Papitto} {et~al.}(2013{\natexlab{a}}){Papitto}, {Ferrigno}, {Bozzo},
  {Rea}, {Pavan}, {Burderi}, {Burgay}, {Campana}, {di Salvo}, {Falanga},
  {Filipovi{\'c}}, {Freire}, {Hessels}, {Possenti}, {Ransom}, {Riggio},
  {Romano}, {Sarkissian}, {Stairs}, {Stella}, {Torres}, {Wieringa}, \&
  {Wong}}]{pfb+13}
{Papitto}, A., {Ferrigno}, C., {Bozzo}, E., {Rea}, N., {Pavan}, L., {Burderi},
  L., {Burgay}, M., {Campana}, S., {di Salvo}, T., {Falanga}, M.,
  {Filipovi{\'c}}, M.~D., {Freire}, P.~C.~C., {Hessels}, J.~W.~T., {Possenti},
  A., {Ransom}, S.~M., {Riggio}, A., {Romano}, P., {Sarkissian}, J.~M.,
  {Stairs}, I.~H., {Stella}, L., {Torres}, D.~F., {Wieringa}, M.~H., \& {Wong},
  G.~F. 2013{\natexlab{a}}, \nat, 501, 517

\bibitem[{{Papitto} {et~al.}(2013{\natexlab{b}}){Papitto}, {Hessels}, {Burgay},
  {Ransom}, {Rea}, {Possenti}, {Stairs}, {Ferrigno}, \& {Bozzo}}]{phb+13}
{Papitto}, A., {Hessels}, J. W.~T., {Burgay}, M., {Ransom}, S., {Rea}, N.,
  {Possenti}, A., {Stairs}, I., {Ferrigno}, C., \& {Bozzo}, A.
  2013{\natexlab{b}}, The Astronomer's Telegram, 5069, 1

\bibitem[{{Patruno} {et~al.}(2013){Patruno}, {Archibald}, {Hessels},
  {Bogdanov}, {Stappers}, {Bassa}, {Janssen}, {Kaspi}, {Tendulkar}, \&
  {Lyne}}]{pah+13}
{Patruno}, A., {Archibald}, A.~M., {Hessels}, J.~W.~T., {Bogdanov}, S.,
  {Stappers}, B.~W., {Bassa}, C.~G., {Janssen}, G.~H., {Kaspi}, V.~M.,
  {Tendulkar}, S., \& {Lyne}, A.~G. 2013, ArXiv e-prints

\bibitem[{{Patruno} {et~al.}(2012){Patruno}, {Bult}, {Gopakumar}, {Hartman},
  {Wijnands}, {van der Klis}, \& {Chakrabarty}}]{pbg+12}
{Patruno}, A., {Bult}, P., {Gopakumar}, A., {Hartman}, J.~M., {Wijnands}, R.,
  {van der Klis}, M., \& {Chakrabarty}, D. 2012, \apjl, 746, L27

\bibitem[{{Pletsch} {et~al.}(2012){Pletsch}, {Guillemot}, {Allen}, {Kramer},
  {Aulbert}, {Fehrmann}, {Ray}, {Barr}, {Belfiore}, {Camilo}, {Caraveo}, {{\c
  C}elik}, {Champion}, {Dormody}, {Eatough}, {Ferrara}, {Freire}, {Hessels},
  {Keith}, {Kerr}, {de Luca}, {Lyne}, {Marelli}, {McLaughlin}, {Parent},
  {Ransom}, {Razzano}, {Reich}, {Saz Parkinson}, {Stappers}, \&
  {Wolff}}]{pga+12}
{Pletsch}, H.~J., {Guillemot}, L., {Allen}, B., {Kramer}, M., {Aulbert}, C.,
  {Fehrmann}, H., {Ray}, P.~S., {Barr}, E.~D., {Belfiore}, A., {Camilo}, F.,
  {Caraveo}, P.~A., {{\c C}elik}, {\"O}., {Champion}, D.~J., {Dormody}, M.,
  {Eatough}, R.~P., {Ferrara}, E.~C., {Freire}, P.~C.~C., {Hessels}, J.~W.~T.,
  {Keith}, M., {Kerr}, M., {de Luca}, A., {Lyne}, A.~G., {Marelli}, M.,
  {McLaughlin}, M.~A., {Parent}, D., {Ransom}, S.~M., {Razzano}, M., {Reich},
  W., {Saz Parkinson}, P.~M., {Stappers}, B.~W., \& {Wolff}, M.~T. 2012, \apj,
  744, 105

\bibitem[{{Ransom}(2001)}]{rans01}
{Ransom}, S.~M. 2001, PhD thesis, Harvard University

\bibitem[{{Ransom} {et~al.}(2009){Ransom}, {Demorest}, {Ford}, {McCullough},
  {Ray}, {DuPlain}, \& {Brandt}}]{rdf+09}
{Ransom}, S.~M., {Demorest}, P., {Ford}, J., {McCullough}, R., {Ray}, J.,
  {DuPlain}, R., \& {Brandt}, P. 2009, in American Astronomical Society Meeting
  Abstracts, Vol. 214, American Astronomical Society Meeting Abstracts \#214,
  605.08

\bibitem[{{Rappaport} {et~al.}(1983){Rappaport}, {Verbunt}, \& {Joss}}]{rvj83}
{Rappaport}, S., {Verbunt}, F., \& {Joss}, P.~C. 1983, \apj, 275, 713

\bibitem[{{Ray} {et~al.}(2013){Ray}, {Ransom}, {Cheung}, {Giroletti},
  {Cognard}, {Camilo}, {Bhattacharyya}, {Roy}, {Romani}, {Ferrara},
  {Guillemot}, {Johnston}, {Keith}, {Kerr}, {Kramer}, {Pletsch}, {Saz
  Parkinson}, \& {Wood}}]{rrc+13}
{Ray}, P.~S., {Ransom}, S.~M., {Cheung}, C.~C., {Giroletti}, M., {Cognard}, I.,
  {Camilo}, F., {Bhattacharyya}, B., {Roy}, J., {Romani}, R.~W., {Ferrara},
  E.~C., {Guillemot}, L., {Johnston}, S., {Keith}, M., {Kerr}, M., {Kramer},
  M., {Pletsch}, H.~J., {Saz Parkinson}, P.~M., \& {Wood}, K.~S. 2013, \apjl,
  763, L13

\bibitem[{{Roberts}(2011)}]{robe11}
{Roberts}, M.~S.~E. 2011, in American Institute of Physics Conference Series,
  Vol. 1357, American Institute of Physics Conference Series, ed. M.~{Burgay},
  N.~{D'Amico}, P.~{Esposito}, A.~{Pellizzoni}, \& A.~{Possenti}, 127--130

\bibitem[{{Sabbi} {et~al.}(2003){Sabbi}, {Gratton}, {Ferraro}, {Bragaglia},
  {Possenti}, {D'Amico}, \& {Camilo}}]{sgf+03}
{Sabbi}, E., {Gratton}, R., {Ferraro}, F.~R., {Bragaglia}, A., {Possenti}, A.,
  {D'Amico}, N., \& {Camilo}, F. 2003, \apjl, 589, L41

\bibitem[{{Shklovskii}(1970)}]{shkl70}
{Shklovskii}, I.~S. 1970, Scientific American, 13, 562

\bibitem[{{Stappers} {et~al.}(2013){Stappers}, {Archibald}, {Bassa}, {Hessels},
  {Janssen}, {Kaspi}, {Lyne}, {Patruno}, \& {Hill}}]{sab+13}
{Stappers}, B.~W., {Archibald}, A., {Bassa}, C., {Hessels}, J., {Janssen}, G.,
  {Kaspi}, V., {Lyne}, A., {Patruno}, A., \& {Hill}, A.~B. 2013, The
  Astronomer's Telegram, 5513, 1

\bibitem[{{Stappers} {et~al.}(2001){Stappers}, {Bailes}, {Lyne}, {Camilo},
  {Manchester}, {Sandhu}, {Toscano}, \& {Bell}}]{sbl+01}
{Stappers}, B.~W., {Bailes}, M., {Lyne}, A.~G., {Camilo}, F., {Manchester},
  R.~N., {Sandhu}, J.~S., {Toscano}, M., \& {Bell}, J.~F. 2001, \mnras, 321,
  576

\bibitem[{{Stappers} {et~al.}(2003){Stappers}, {Gaensler}, {Kaspi}, {van der
  Klis}, \& {Lewin}}]{sgk+03}
{Stappers}, B.~W., {Gaensler}, B.~M., {Kaspi}, V.~M., {van der Klis}, M., \&
  {Lewin}, W.~H.~G. 2003, Science, 299, 1372

\bibitem[{{Takata} {et~al.}(2012){Takata}, {Cheng}, \& {Taam}}]{tct12}
{Takata}, J., {Cheng}, K.~S., \& {Taam}, R.~E. 2012, \apj, 745, 100

\bibitem[{{Tam} {et~al.}(2010){Tam}, {Hui}, {Huang}, {Kong}, {Takata}, {Lin},
  {Yang}, {Cheng}, \& {Taam}}]{thh+10}
{Tam}, P.~H.~T., {Hui}, C.~Y., {Huang}, R.~H.~H., {Kong}, A.~K.~H., {Takata},
  J., {Lin}, L.~C.~C., {Yang}, Y.~J., {Cheng}, K.~S., \& {Taam}, R.~E. 2010,
  \apjl, 724, L207

\bibitem[{{Thorstensen} \& {Armstrong}(2005)}]{ta05}
{Thorstensen}, J.~R. \& {Armstrong}, E. 2005, \aj, 130, 759

\bibitem[{{Vink} {et~al.}(2011){Vink}, {Bamba}, \& {Yamazaki}}]{vby11}
{Vink}, J., {Bamba}, A., \& {Yamazaki}, R. 2011, \apj, 727, 131

\bibitem[{{Wang} {et~al.}(2013){Wang}, {Wang}, \& {Morrell}}]{wwm13}
{Wang}, X., {Wang}, Z., \& {Morrell}, N. 2013, \apj, 764, 144

\end{thebibliography}

\begin{table}
    \caption{Long-term ephemeris for J1023}\label{tab:longterm}
\centering
\begin{tabular}{ll}
\hline\hline
\multicolumn{2}{c}{Fit and data-set} \\
\hline
Pulsar name\dotfill & J1023+0038 \\ 
MJD range\dotfill & 54766.5---56146.6 \\ 
Number of TOAs\dotfill & 7478 \\
Rms timing residual ($\mu s$)\dotfill & 114.0 \\
Weighted fit\dotfill &  No \\ 
\hline
\multicolumn{2}{c}{Measured Quantities} \\ 
\hline
Pulse frequency, $\nu$ (s$^{-1}$)\dotfill & 592.42145906986(10) \\ 
First derivative of pulse frequency, $\dot{\nu}$ (s$^{-2}$)\dotfill & $-$2.432(3)$\times 10^{-15}$ \\ 
Orbital period, $P_b$ (d)\dotfill & 0.1980963569(3) \\ 
Epoch of periastron, $T_0$ (MJD)\dotfill & 54905.9713992(3) \\ 
Projected semi-major axis of orbit, $x$ (lt-s)\dotfill & 0.343343(3) \\ 
First derivative of orbital period, $\dot{P_b}$\dotfill & $-$7.32(6)$\times 10^{-11}$ \\ 
\hline
\multicolumn{2}{c}{Set Quantities} \\ 
\hline
Right ascension, $\alpha$\dotfill & 10:23:47.687198 \\ 
Declination, $\delta$\dotfill & +00:38:40.84551 \\ 
Epoch of frequency determination (MJD)\dotfill & 54906 \\ 
Epoch of position determination (MJD)\dotfill & 54995 \\ 
Dispersion measure, $DM$ (cm$^{-3}$pc)\dotfill & 14.3308 \\ 
Proper motion in right ascension, $\mu_{\alpha}$ (mas\,yr$^{-1}$)\dotfill & 4.76 \\ 
Proper motion in declination, $\mu_{\delta}$ (mas\,yr$^{-1}$)\dotfill & $-$17.34 \\ 
Parallax, $\pi$ (mas)\dotfill & 0.000731 \\ 
Orbital eccentricity, $e$\dotfill & 0 \\ 
\hline
\end{tabular}

Note: Figures in parentheses are  the nominal 1$\sigma$ \textsc{tempo2} uncertainties in the least-significant digits quoted. Time units are in barycentric coordinate time (TCB). 
\end{table}

\end{document}